\documentclass[referee]{raa}            

\usepackage{graphicx,times}             
\usepackage{natbib}
\usepackage{amssymb,amsmath}
\bibpunct{(}{)}{;}{a}{}{,}

\usepackage[pagebackref=true]{hyperref}

\begin{document}
\title{The Next-Generation 21CMA Telescope: Design, Commissioning, and Instrumental Effects in an SKA-LFAA-Like System}
\titlerunning{The Next-Generation 21CMA}
\authorrunning{J.-H. Gu et al.}
\author{Jun-Hua Gu\inst{1,*} \footnotetext{$*$ Corresponding Authors, these authors contributed equally to this work.}, Quan Guo\inst{2}, Liang Dong\inst{3}, Yu-Kai Zhou\inst{4}, Kuan-Jun Li\inst{1}, Yan Huang\inst{1}, Jing-Ying Wang\inst{2}, Wei-Wei Zhu\inst{1}, You-Ling Yue\inst{1}, Rui Cao\inst{5}, Guo-Liang Peng\inst{5}, Xiao-Hui Tao\inst{5}, Li-Hui Jiang\inst{5}, Ya-Jun Wu\inst{2} }
\institute{State Key Laboratory of Radio Astronomy and Technology, National Astronomical Observatories,
	Chinese Academy of Sciences, 20A Datun Road, Beijing 100101, China\\
	\and
	State Key Laboratory of Radio Astronomy and Technology, Shanghai Astronomical Observatory,
	Chinese Academy of Sciences, 80 Nandan Road, Shanghai 200030, China
	\and
	Yunnan Astronomical Observatory, Chinese Academy of Sciences, Yunnan, 650216, China
	\and
	Department of Astronomy, School of Physics, Peking University, Beijing 100871, China
	\and
	Key Laboratory of Aperture Array and Space Application, Hefei 230088, China
}

\abstract{As the Square Kilometre Array (SKA) approaches operational status, its complex digital architecture introduces new instrumental challenges. To explore relevant observational and data processing strategies, we have upgraded the 21CMA telescope to the Next-Generation 21CMA (Ng21CMA). This paper presents the design and commissioning of the Ng21CMA system, featuring a digital backend capable of real-time beamforming. We demonstrate its performance through interferometric observations and high-time-resolution pulsar measurements, validating the system's sensitivity and operational stability. As a representative example of instrumental effects accessible with this platform, we investigate the impact of the two-stage channelization strategy used in SKA-LFAA-like systems. We show that it introduces a sawtooth-like spectral structure (SLOSS), characterized using both simulations and observational data. These results provide useful references for understanding instrument-induced spectral features and for guiding system design and calibration in future large-scale aperture arrays.
	\keywords{methods:observational --- cosmology: observations, dark ages, reionization --- techniques:interferometric --- radio continuum:general} 
}

\maketitle

\section{Introduction}
The Square Kilometre Array (SKA) \footnote{\url{https://www.skao.int}} 
is set to generate scientific data in the near future. 
One of its key components is the Low-Frequency Aperture Array (LFAA), which operates in the $50-350$ MHz frequency range. 
A critical scientific objective of the LFAA is the detection of the neutral hydrogen 21 cm line signal from the Epoch of Reionisation (EoR). The brightness temperature of this neutral hydrogen 21 cm line is at least four orders of magnitude weaker than the intense foreground continuum emission \citep{2006PhR...433..181F}, which primarily originates from the Milky Way and extragalactic radio sources. This significant contrast between the desired signal and the foreground emission presents substantial challenges for both the instrumentation and the data reduction pipelines. Achieving measurement and calibration precision better than $10^{-4}$ is therefore essential for the successful detection of the EoR signals.

Given that the LFAA is one of the most complex astronomical instruments ever developed, the presence of instrumental effects is inevitable and must be anticipated. As a result, methods need to be devised to mitigate any potential effects that may arise. Among the advanced technologies integrated into the LFAA, digital beamforming (DBF) plays a pivotal role in meeting the scientific demands of the SKA \citep[e.g.,][]{2015PhDT.......574S, 2010arXiv1008.4051K}. In the meter-wave radio band, traditional mechanical structures for steering the antenna beam are no longer feasible. Instead, digital and electronic techniques are employed to compensate for geometric delay, enabling the antenna beam to be steered electronically in multiple directions.

When wide-band signal processing is implemented in large-scale systems like the SKA, where the ratio $(\nu_{\max}-\nu_{\min})/(\nu_{\max}+\nu_{\min})$ exceeds $1/2$ (with $\nu_{\min}$ and $\nu_{\max}$ denoting the lower and upper bounds of the working band), the real-time computational demands of digital beamforming become extremely high \citep{2010arXiv1008.4051K}. This leads to power consumption and data transfer bandwidth approaching the system’s upper limits. Consequently, a dilemma arises between ensuring efficient system cooling and maintaining adequate radio frequency interference (RFI) shielding: a sufficiently well-shielded enclosure may impede cooling, while effective cooling could compromise RFI shielding.

\begin{figure}[hbt!]
	\centering
	\includegraphics[width=\columnwidth]{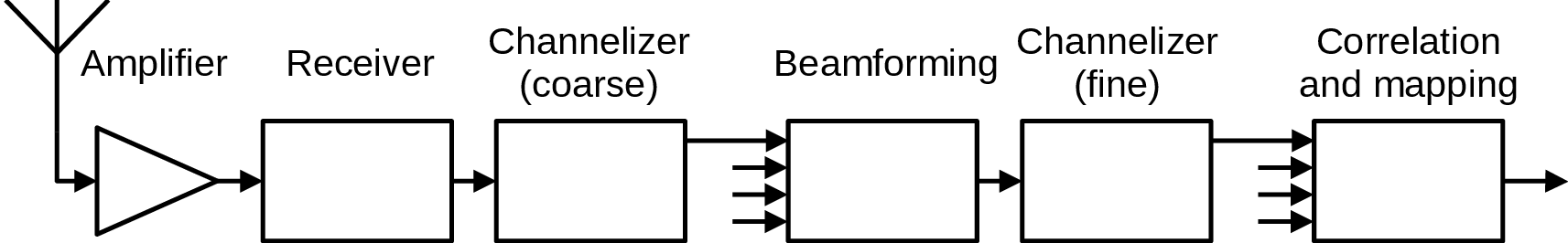}
	\caption{\label{fig:flow_chart} The data flow chart of the LFAA data acquisition system.
		This chart was initially presented in \citet{2017MmSAI..88..154C}, and we have reproduced it here for reference.
	}
\end{figure}

To address these challenges, various engineering compromises must be made. One such compromise is the adoption of a two-stage channelization strategy. This design is implemented in the SKA-LFAA signal processing chain \citep[e.g.,][]{2017JAI.....641015C, 2020SPIE11450E..03C}. The signal processing flowchart is shown in Figure \ref{fig:flow_chart}. In the station beamformer, the received radio frequency signal undergoes initial channelization to produce a series of coarse channels, each with a bandwidth of 781.25 kHz. Digital beamforming is then applied to these coarse channels. The data stream for each beam at each station is subsequently routed to a central correlator, where a second channelization step is applied to generate fine channels tailored to the specific scientific objectives. By avoiding direct channelization of the received signal to the final frequency resolution, computational demands at the station level are alleviated, enabling the LFAA to operate within technological and budgetary constraints.
It should be noted that while the two-stage channelization strategy in the SKA-LFAA shares a conceptual lineage with earlier digital receiver architectures such as that of Murchison Widefield Array \citep[MWA][]{2013PASA...30....7T}, their functional roles differ significantly. In the MWA architecture \citep{2015ExA....39...73P}, the primary role of the initial coarse channelization is to partition the total bandwidth into a series of selectable channels for downstream processing, following an analog beamforming stage. In contrast, the SKA-LFAA strategy performs the first-stage channelization prior to digital beamforming. By applying DBF at the coarse-channel level rather than the full-bandwidth raw data stream, the system significantly reduces the sampling rate and computational load required for phase-shifting and summation, thereby enabling more complex and flexible beamforming configurations within the available hardware resources.

In principle, frequency-domain beamforming based on phase weighting, as adopted in the SKA-LFAA, introduces the well-known “beam squint” effect \citep[e.g.,][]{garakoui2011phased} when the signal bandwidth is finite. In systems employing two-stage channelization, this effect becomes structured in frequency, giving rise to the observed sawtooth-like spectral features, which will be discussed in detail in the following section. 
A study by \citet{2016PASA...33...19T} identified non-homogeneity among fine channels within the same coarse channel and proposed calibration algorithms for this effect.
The two-stage channelization strategy has also been adopted in The LOw-Frequency ARray \citep[LOFAR][]{2013A&A...556A...2V}. In that system, bandpass non-uniformities introduced by the channelization filters are corrected after the second-stage channelization. However, the impact of frequency-dependent beamforming effects, such as beam squint, on the resulting spectral structure has not been explicitly addressed in previous studies to the best of our knowledge.
Consistent with this, the NenuFAR telescope \citep[New extension in Nan\c{c}ay upgrading LOFAR][]{zarka2020low}, which shares part of its signal processing architecture with LOFAR, has reported similar sawtooth-like variations at the level of approximately $0.1$–$0.2$ dB in beamformed observations (NenuFAR Astronomers Page\footnote{\url{https://nenufar.obs-nancay.fr/en/astronomer/}}).
However, a comprehensive evaluation of these instrumental effects, particularly in the context of EoR detection, is still lacking.

Given that the neutral hydrogen 21 cm line signal from the EoR is much weaker than the foreground emission, achieving precise calibration is critical. The impact of the two-stage channelization strategy on EoR detection remains underexplored, making it essential for the community to address this gap.
In addition to the effects introduced by the two-stage channelization strategy, the LFAA system is also subject to other instrumental effects, such as beamforming errors and mutual coupling between antennas. While these effects are not discussed in this paper, they are also significant factors to consider in the performance of the LFAA system.

To study the instrumental effects anticipated in future SKA scientific data, we have upgraded the 21 Centimeter Array telescope \citep{2005ASPC..345..441P, 2016RAA....16...36H, 2016ApJ...832..190Z} to the next-generation 21CMA (Ng21CMA) telescope. The Ng21CMA telescope reuses the antenna stations of the original 21CMA telescope and has been equipped with a new data acquisition system capable of digital beamforming.

This paper presents the system design of the Ng21CMA in Section \ref{sec:ng21cma}, followed by preliminary observational results in Section \ref{sec:obs}. Section \ref{sec:explaination} includes numerical simulations of the spectral response of the station-level DBF subsystem, examines the cause of the sawtooth-like spectral structure, and proposes a calibration method. A brief discussion is provided in Section \ref{sec:discussion}, and the paper concludes in Section \ref{sec:conclusions}. 

These results provide new insight into frequency-dependent instrumental effects in two-stage channelization beamforming systems, which have not been systematically characterized in previous studies.

\section{The Next-Generation 21CMA Digital Beamforming System}
\label{sec:ng21cma}
\subsection{System Overview}

\begin{figure}[hbt!]
	\centering
	\includegraphics[width=0.9\columnwidth]{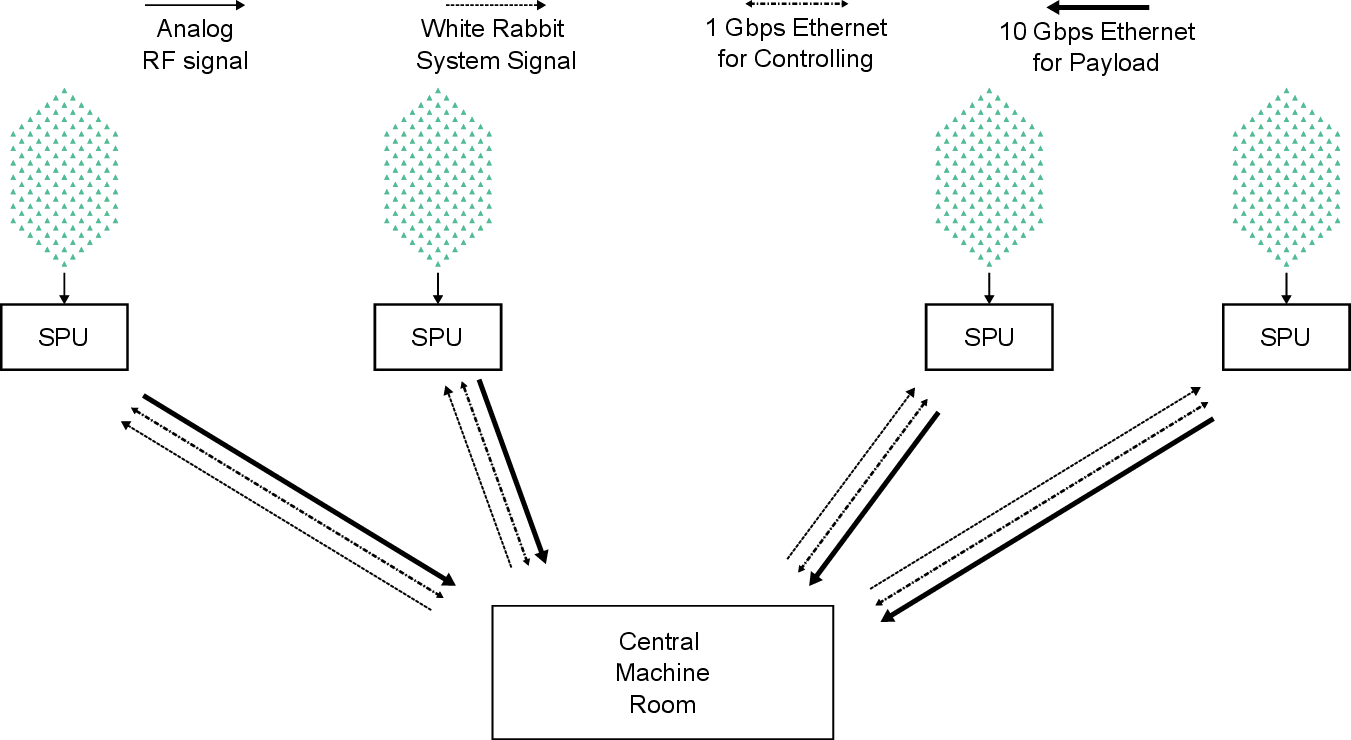}
	\caption{\label{fig:ng21cma_block_diagram} The block diagram of the Ng21CMA data acquisition system.}
\end{figure}

Building upon the original 21CMA antenna stations, we are upgrading the system with digital beamforming technology. A new data acquisition system has been installed, and for testing purposes, an ad-hoc GPU-based software correlator has been set up. The block diagram of the Ng21CMA data acquisition system is shown in Figure \ref{fig:ng21cma_block_diagram}. The process begins when the radio frequency signal, received by each 21CMA log-periodic antenna, is first amplified by a low-noise amplifier (LNA). The amplified signal is then transmitted via coaxial cables to the station processing unit (SPU).
Within the SPU, the signal is further amplified to ensure it meets the dynamic range requirements of the analog-to-digital converters (ADC), and subsequently digitized. The digitized time-domain signals are channelized using a set of polyphase filterbanks (PFB), which are then utilized to form the station beams.
Finally, the data streams from each beam of each station are transmitted to the central correlator located in the machine room, where they are processed to produce the visibility results.

\subsection{RF Analog Sub-system}
\begin{figure}[hbt!]
	\centering
	\includegraphics[width=0.9\columnwidth]{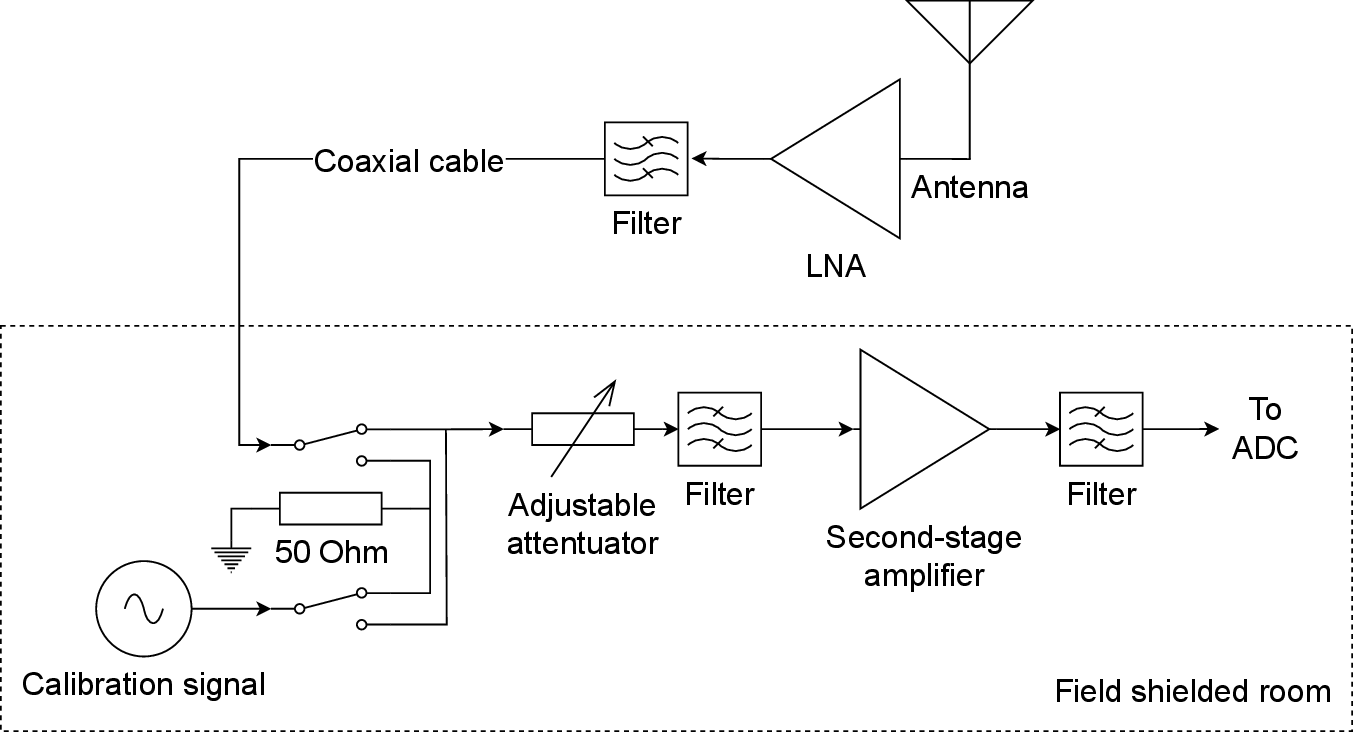}
	\caption{\label{fig:ng21cma_rf_sys} The RF analog sub-system behind each antenna of Ng21CMA.}
\end{figure}

The radio frequency (RF) analog subsystem, depicted in Figure \ref{fig:ng21cma_rf_sys}, includes a set of low-noise amplifiers (LNAs) integrated into all 21CMA log-periodic antennas, along with a second-stage amplifier. Each RF chain behind the antennas is equipped with an adjustable attenuator (up to 31 dB) to fine-tune the overall gain of the RF chain. The passband of the RF filters is designed for the 50–200 MHz frequency range, which defines the operating bandwidth of the Ng21CMA system.
A self-calibration path is incorporated into the RF analog subsystem. When required, the RF switch activates calibration mode, routing a calibration signal (a single-tone sine wave with an adjustable frequency) into the second-stage amplifier instead of the signal received by the antenna. By varying the frequency of the calibration signal, we can calibrate the relative delay and amplitude response differences between different RF chains.
Amplitude response calibration is straightforward, while the calibration of relative delay differences between any two RF chains is performed using the phase-frequency relationship of the cross-correlation results. Specifically, the delay between a selected pair of RF chains, denoted as $p$ and $q$, is calculated by solving the least-squares fitting problem

\begin{equation}
	\Delta\tau_{pq}=\arg \min_{\Delta \tau_{pq}} \sum_j \left |\frac{<x_{p}(\nu_j) {x}^*_q(\nu_j)>}{<|x_{p}|><|x_{q}|>}-\exp\left( i 2\pi \Delta\tau_{pq} \nu_j\right)\right |^2,
\end{equation}
where $x(\nu_j)$'s denote the channelized signal of the $j$-th coarse channel, $x^*$ denotes the complex conjugation, $<\cdot>$ denotes averaging.

\subsection{The Station Configuration}
\label{ssec:station}
The Ng21CMA system reuses the antenna stations from the original 21CMA telescope. Each station consists of 127 single-polarized log-periodic antennas arranged in a hexagonal pattern, all pointing toward the north celestial pole. In the original 21CMA telescope, the station beam was directed toward the north celestial pole by adjusting the lengths of the coaxial cables. The signals received by each antenna were combined using an RF combiner and transmitted via an analog fiber link to the central data acquisition and processing facility.

In contrast, the upgraded Ng21CMA system employs digital beamforming technology to electronically steer the station beam. By adjusting the phase delays of the digital signals from each antenna, the station beam can be directed to any desired location within the station's field of view. The beam can be steered in real time to track specific targets or to scan the sky.

\begin{figure}[hbt!]
	\centering
	\includegraphics[width=1\columnwidth]{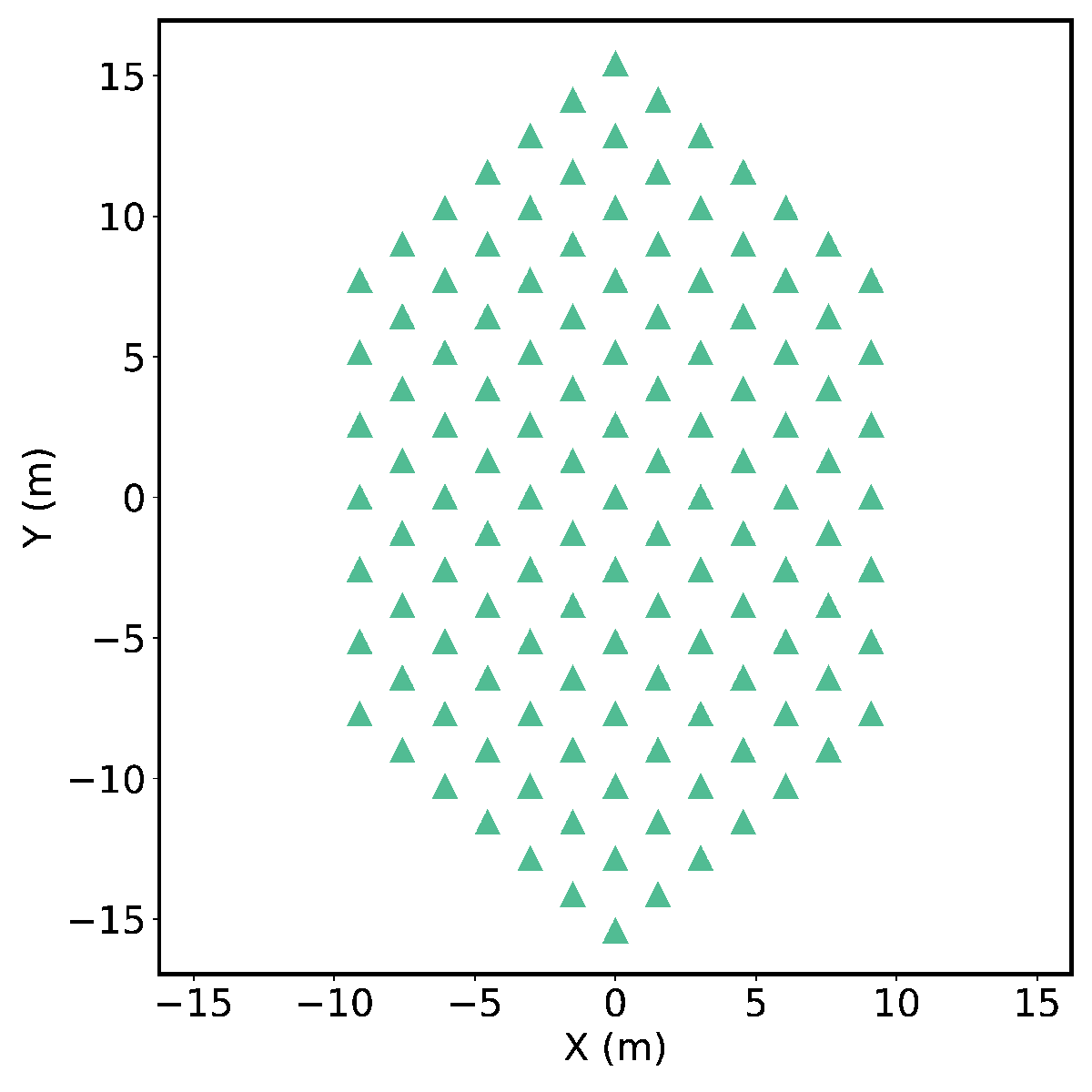}
	\caption{\label{fig:21cma_station} The station configuration of 21CMA.}
\end{figure}

The configuration of each 21CMA station is identical, as shown in Figure \ref{fig:21cma_station}. The station beam pattern of the original 21CMA system was optimized for pointing toward the north celestial pole, resulting in an elongated configuration along the north-south direction.

\subsection{Digital Beamformer Unit}
The digital beamformer unit (DBU) is implemented using an FPGA architecture, with a \textit{PowerPC} single-board computer serving as the controller. Each station's DBU features 128 data acquisition ports, 127 of which are connected to the RF chains behind each antenna. The beamformer operates using a frequency-domain algorithm, similar to that employed in the SKA-LFAA \citep{2017JAI.....641015C}. For each coarse channel across all data acquisition ports, complex weights are applied and summed to form the desired beam.
The complex channelized station output is calculated as
\begin{equation}
	S(t, c)=\sum_a \exp[2\pi i \nu_c \tau(t, a)] \tilde{U}(t, \nu_c, a),
\end{equation}
where $t$ denotes the time, $c$ denotes the channel index, $a$ denotes the antenna index, $\nu_c$ is the central frequency of the $c$-th coarse spectral channel, and $\tau$ is the desired time-domain delay.
There are two differences between our implementation and that of the SKA-LFAA: (1) the desired time-domain delay is updated at fixed intervals, instead of being continuously updated as in the SKA-LFAA; (2) the oversampling factor is set to $2$, rather than the $32/27$ used by the SKA-LFAA.
The oversampling factor of $2$ makes it more convenient to design the second-stage channelizer.
Two sets of complex weights can be configured, allowing two beams to be formed simultaneously for each coarse channel.
The resulting payload data stream is then sent to the central machine room for further processing.

\subsection{Inter-station Clock Synchronization}
The data acquisition for each station is performed locally, right next to the antennas.
To ensure phase delay stabilization among all stations, the sampling clock must be synchronized, and the wall-clock time of the first data point of each observation must be aligned.
We use the so-called \textit{White Rabbit} \citep[WR, ][]{5340196}  system to broadcast the 10 MHz reference clock and the 1-PPS (one pulse per second) signal from the central machine room to all the involved stations.
The WR system provides sub-nanosecond timing precision among all the slave nodes, meeting the requirements for observations in the meter-wave band.

\subsection{Ethernet-based Monitoring and Controlling Network}
All the SPUs and a controlling computer in the central machine room are connected to a 1 Gbps Ethernet local area network.
The instructions for controlling the data acquisition tasks are wrapped in UDP (User Datagram Protocol) packets and sent from the controlling computer.
Similarly, device state reports are also wrapped in UDP packets and sent from each SPU.

\subsection{Payload Data Stream}
The payload data streams, consisting of complex channelized and beam-formed data, are transferred from each SPU to the correlator located in the central machine room through a 10 Gbps Ethernet connection.
Jumbo UDP packets are used to wrap the payload data.
The sampling rate is set to 480 MSps, corresponding to a bandwidth of 240 MHz, with time-domain signals channelized into 512 coarse channels.
For each SPU, we use a set of four pairs of fibers to transfer the payload data, with each pair transferring 100 coarse channels, corresponding to a bandwidth of 46.875 MHz.
The effective data rate on each pair of fiber is 375 $\rm{MByte~s^{-1}}$.

\subsection{The Second Stage Channelizer and Software Correlator}
In the central machine room, each coarse channel is further channelized into $n$ fine channels (where $n \geq 2$ is an integer) using a set of critically sampled PFBs implemented with GPUs (Graphics Processing Units). Of these, $m = n(f-1)/f$ fine channels are discarded in the overlapping frequency ranges near the channel edges to implement the two-stage channelization strategy.
For an oversampling factor of $f=2$, the only constraint on $n$ is that it must be an even integer so that the resulting $m$ is an integer.
On the other hand, if the oversampling factor $f$ is chosen to be $32/27$, $n$ must be an integer multiple of $32$ in a straightforward implementation.
The fine channel data streams are then directed into the software correlator implemented with GPUs.

\begin{figure}[hbt!]
	\centering
	\includegraphics[width=0.9\columnwidth]{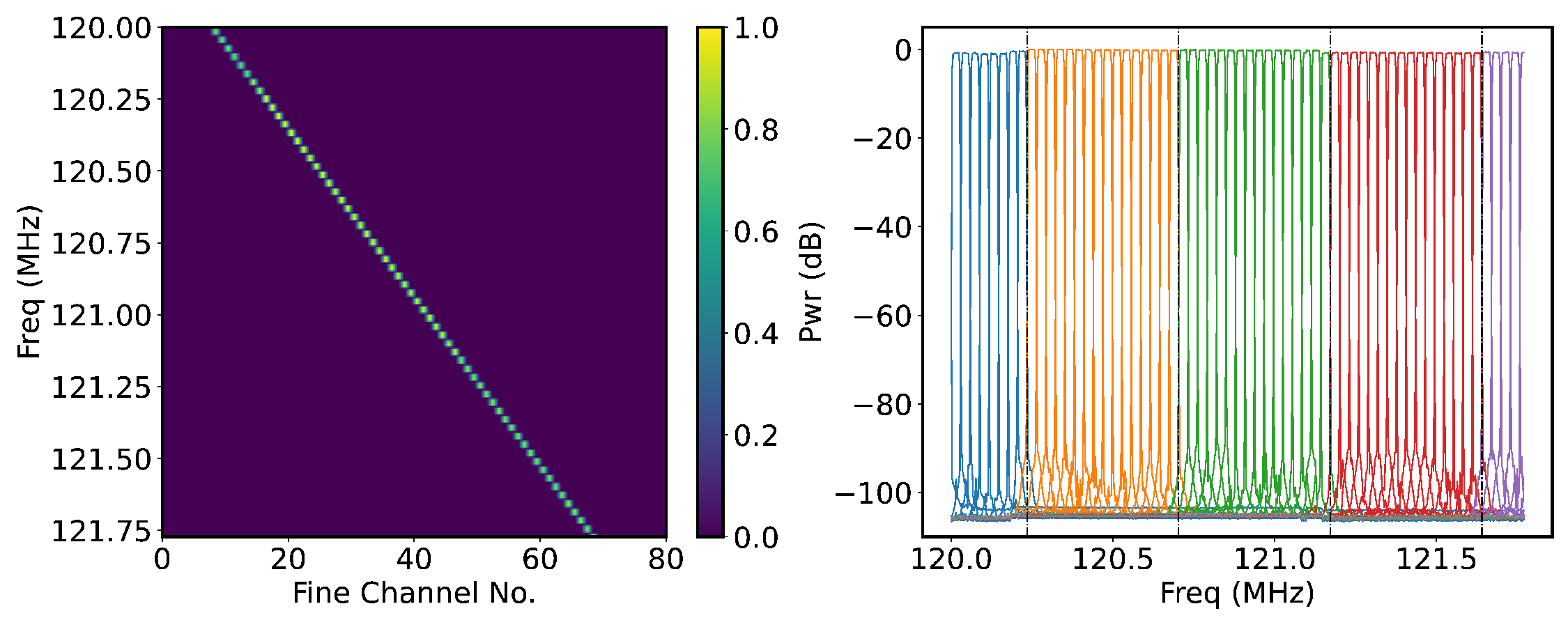}
	\caption{\label{fig:fine_ch_resp} Left: the amplitude of the spectral response matrix of a set of fine channels; Right: the passband profile of a set of fine channels. Fine channels that belong to the same coarse channel are coded in the same color, and the vertical dashed lines mark the coarse channel boundaries.}
\end{figure}

During laboratory tests, the spectral response of the two-stage channelizer is measured with an adjustable single-tone RF signal generator.
The signal generator is connected to the input of the RF chain and configured to emit a set of single-tone signals within the passband of the receiver. For each frequency, the amplitude of the auto-correlation result is recorded.
The amplitude of the spectral response matrix and the passband profile of fine channels are shown in Figure \ref{fig:fine_ch_resp}.

\section{Preliminary Observational Results}
\label{sec:obs}
\subsection{Beam-steering and Target Tracking Validation}
\begin{figure}[hbt!]
	\centering
	\includegraphics[width=0.45\columnwidth]{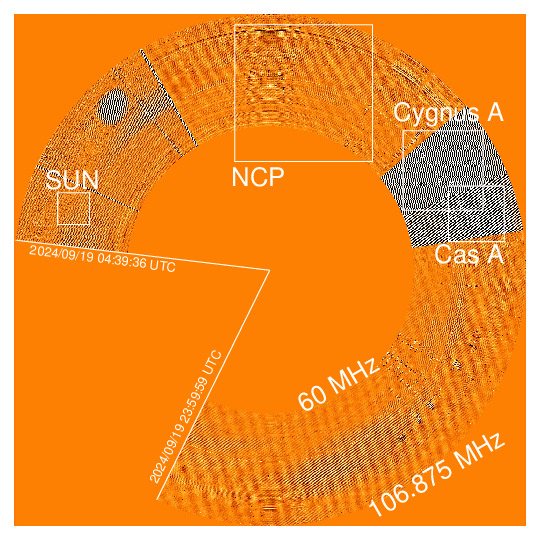}
	\includegraphics[width=0.45\columnwidth]{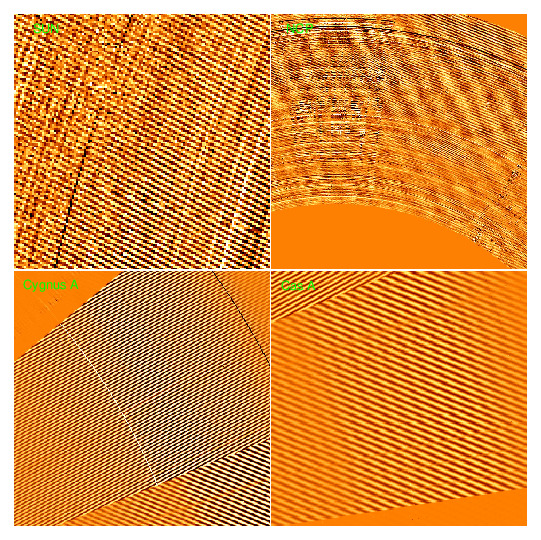}
	\caption{\label{fig:fringes_tracking} On 2024-09-19, we point one formed beam to four targets, namely the sun, the north celestial pole, Cas A, and Cygnus A, and recorded the visibility. Stable fringes were observed. We show the real part of the visibility data of this day here. Left: The whole data; Right: The fringes of the four targets are amplified. Note that the scales are adjusted to show the fringes more clearly.}
\end{figure}

On 2024-09-19, we selected the Sun, the North Celestial Pole (NCP), Cygnus A, and Cas A as the four targets to be tracked.
The real part of the visibility for one baseline and different spectral channels from this day is shown in Figure \ref{fig:fringes_tracking}.
The visibility is shown in polar coordinates, where the polar angle represents the sidereal time and the radius represents the spectral channels.
Stable and clear fringes can be observed.

\begin{figure*}[hbt!]
	\centering
	\includegraphics[width=0.8\textwidth]{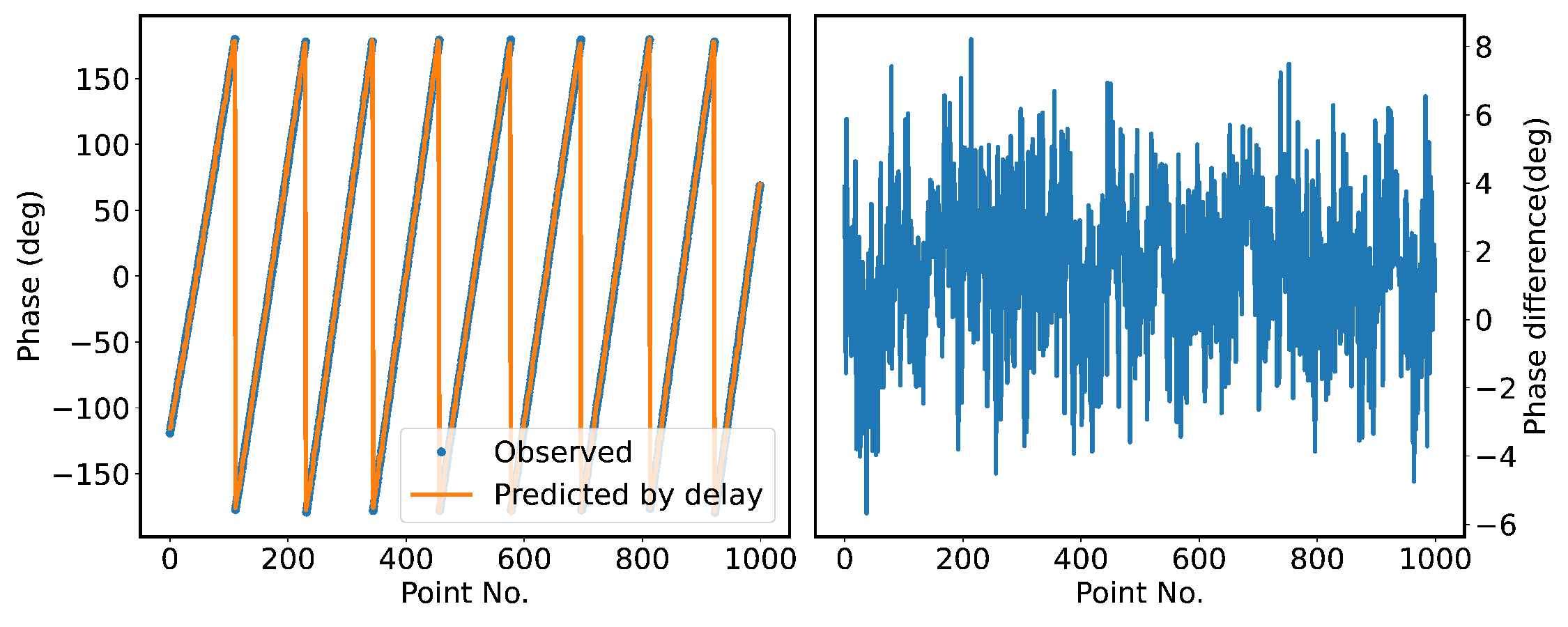}
	\caption{\label{fig:phase_comp} Quantitative phase stability analysis for a representative baseline during a tracking observation of Cas A at 80 MHz. Left: Comparison between the measured visibility phase (blue dots) and the theoretical phase predicted by Eq. \ref{eq:phase_freq} (yellow line), where the delay was determined via a wideband phase-frequency fit. Right: Residual phase difference between the observations and the model, demonstrating a stability of better than $10^\circ$ throughout the integration period.}
\end{figure*}

	Beyond qualitative visual inspection, the phase stability of the DBF system was quantitatively evaluated through a tracking observation of Cas A. The stability was assessed by comparing the measured visibility phase with a theoretical prediction based on the geometric delay difference between stations. Specifically, the delay parameter $\Delta L$ was determined by performing a linear fit to the phase-frequency relationship:
	\begin{equation}
		\phi(\nu)=2\pi \frac{\Delta L}{\lambda}=2\pi \frac{\Delta L}{c}\nu.\label{eq:phase_freq}
	\end{equation}
	where $c$ is the speed of light. The residuals between the observed visibility phase and the model prediction at 80 MHz are presented in Figure \ref{fig:phase_comp}. The measured phase stability is better than $10^\circ$ throughout the observation period, demonstrating that the digital signal chain maintains high coherence and robust timing synchronization, which are essential for high-fidelity interferometric imaging.

\subsection{High Time-resolution Mode for Pulsar Observations}
On 2024-10-28, we observed the PSR B0329+54 pulsar using the high time-resolution mode.
The time resolution is set to be 10 ms, and the beam is set to trace the PSR B0329+54 pulsar. We tried three frequency ranges: $[60, 106.875]$ MHz, [120, 166.875] MHz, and [160, 206.875] MHz.
The observation time is about one hour for each frequency range.
All the three observations show clear pulsar signals after folded.

\begin{figure*}
	\begin{center}
		\includegraphics[width=0.5\textwidth]{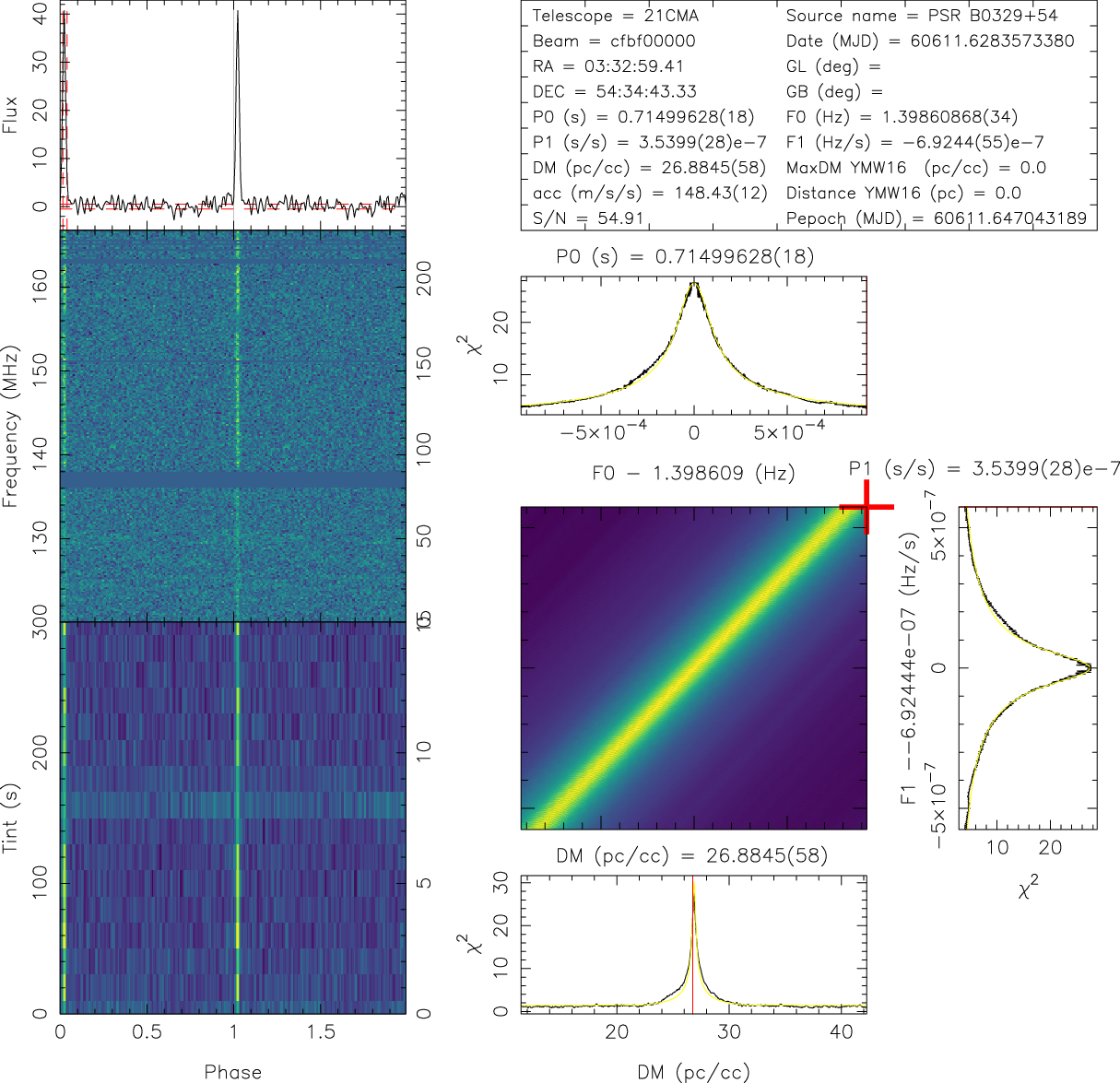}
	\end{center}
	\caption{\label{fig:pulsar} The detection result of the PSR B0329+54 pulsar in the [120, 166.875] MHz sub-band. The observation time is about 5 minutes.}
\end{figure*}

For the [120, 166.875] MHz sub-band, we also performed a 5-minute observation and achieved a statistically significant detection.
The detection result is shown in Figure \ref{fig:pulsar}

\subsection{Drift Scanning and Artificial Spectral Structures Introduced by the Two-Stage Channelization}
\begin{figure}[hbt!]
	\centering
	\includegraphics[width=0.5\columnwidth]{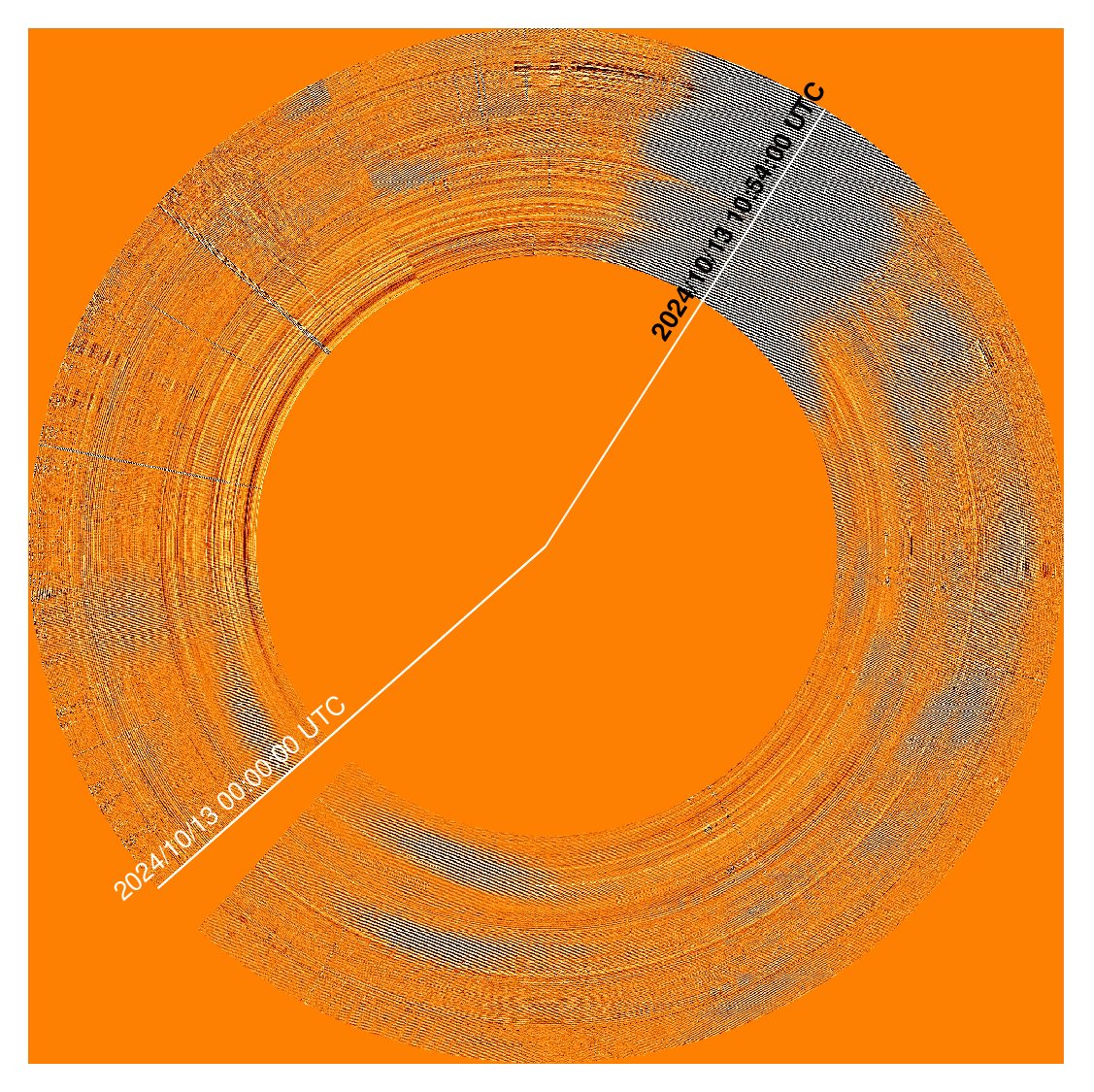}
	\caption{\label{fig:fringes_drift} The real part of the visibility from the observation on 2024-10-13. On this day, after 07:00 UTC, the beam was directed at an $\mathrm{Az}$ of $41.9^\circ$ and an $\mathrm{El}$ of $40.8^\circ$, with Cas A passing through this direction at 10:54 UTC. The coordinate system is the same as that in Figure \ref{fig:fringes_tracking}.}
\end{figure}

\begin{figure}[hbt!]
	\centering
	\includegraphics[width=0.75\columnwidth]{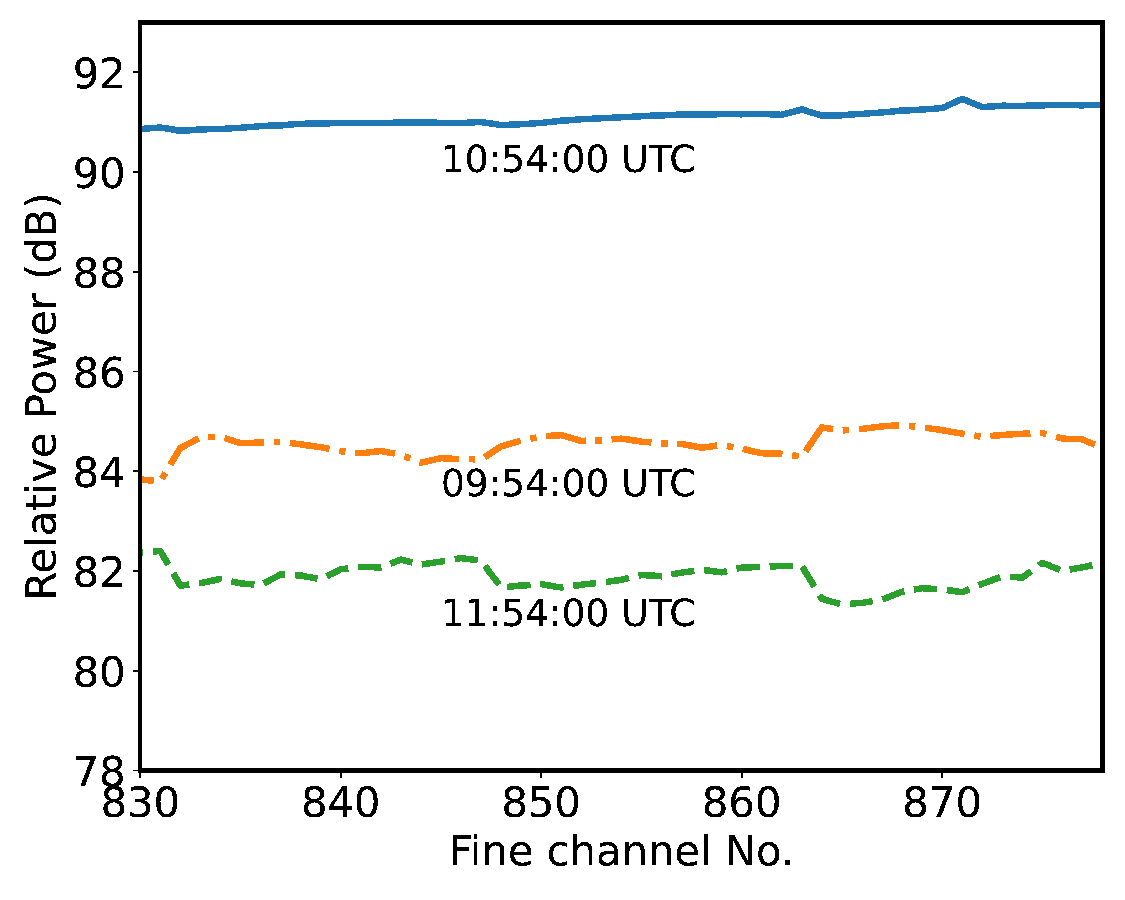}
	\caption{\label{fig:casa_spectrum} This figure shows the raw amplitude of different fine channels of the cross-correlation result from the drift observation on 2024-10-13 at 10:54 UTC, when Cas A passes through the beam center, as well as the amplitude of different fine channels one hour earlier and one hour later. The beam was pointed at an $\mathrm{Az}$ of $41.9^\circ$ and an $\mathrm{El}$ of $40.8^\circ$.}
\end{figure}

Before the instrument was fully developed, we simulated the spectral response of the beamforming system for a radio source offset from the beam center.
We observed that if the beam is not directed at the zenith, the two-stage channelization strategy causes sawtooth-like oscillating spectral structures (hereafter referred to as SLOSS; see Section \ref{sec:explaination}).
To test this hypothesis, on 2024-10-13 after 07:00 UTC, we fixed the beam at an azimuth angle ($\mathrm{Az}$) of $41.9^\circ$ and an elevation angle ($\mathrm{El}$) of $40.8^\circ$, waiting for Cas A to pass through the beam center at 10:54 UTC.
The azimuth angle is defined such that the north is $0^\circ$ and the east is $90^\circ$.
The obtained visibility is shown in Figure \ref{fig:fringes_drift}.
In Figure \ref{fig:casa_spectrum}, we show the amplitude of the raw cross-correlation results for different fine channels at 10:54 UTC, 9:54 UTC, and 11:54 UTC, which can be regarded as an uncalibrated spectrum.
We use cross-correlation results rather than auto-correlation to test our hypothesis because cross-correlation is much less sensitive to noise and radio frequency interference (RFI).
It is evident that at 10:54 UTC, when Cas A is at the beam center, the spectrum is smooth, whereas at 9:54 UTC and 11:54 UTC, the observed spectrum shows a clear sawtooth-like oscillating structure.
This oscillating pattern repeats every 16 fine channels, corresponding to one coarse channel, which can be naturally explained by the two-stage channelization strategy.

\begin{figure}[hbt!]
	\centering
	\includegraphics[width=0.5\columnwidth]{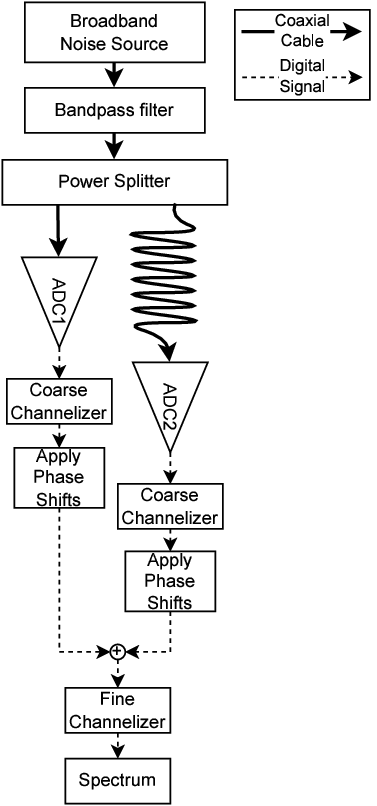}
	\caption{\label{fig:signal_path} Configuration of the digital beamformer used to perform the actual measurement.}
\end{figure}

\begin{figure}[hbt!]
	\centering
	\includegraphics[width=0.9\columnwidth]{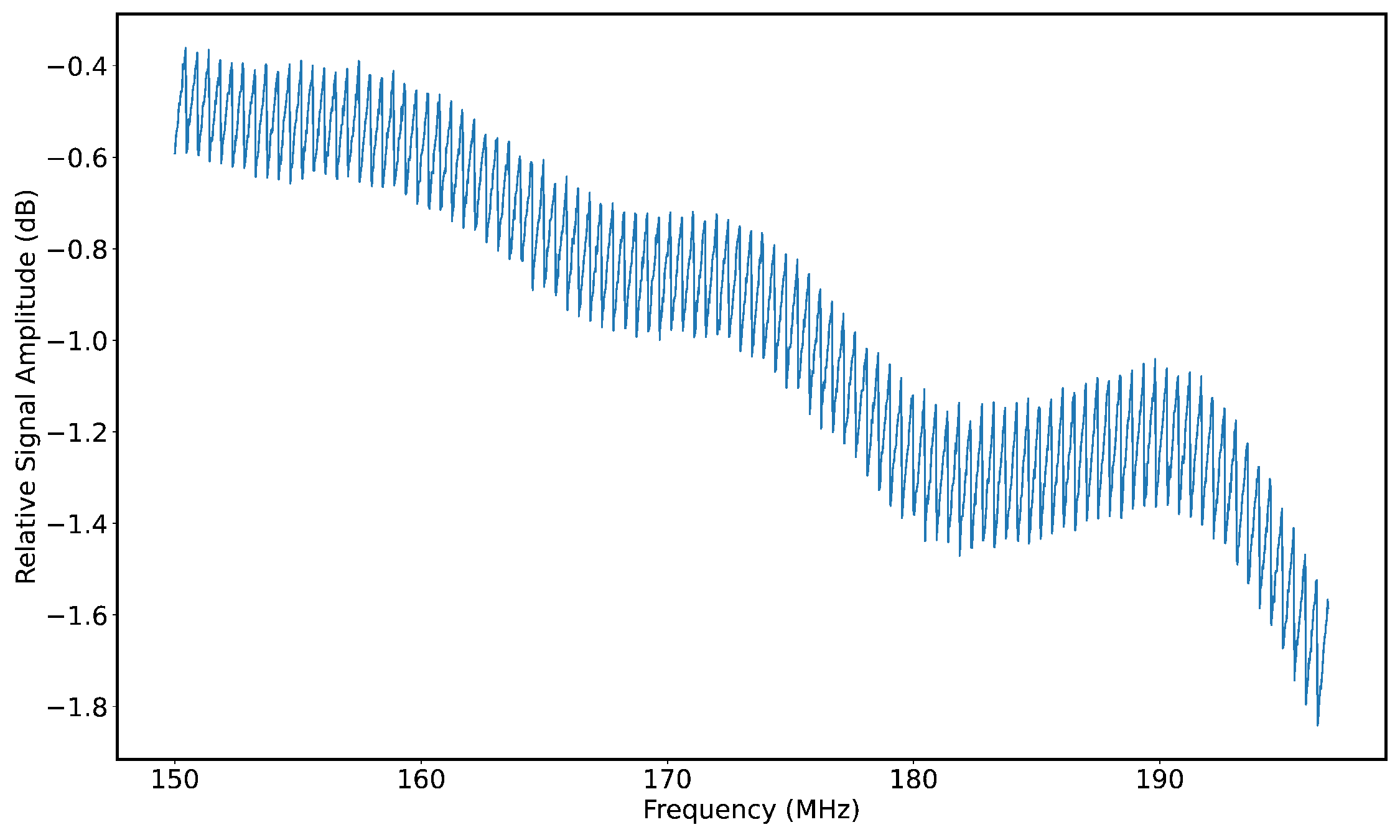}
	\caption{\label{fig:actual_data} The spectral response of an actual beamformer to a broadband noise source, mimicking a hypothetical radio source away from the phase center.}
\end{figure}

The sawtooth-like spectral structure was also observed during laboratory testing of the data acquisition module.
Figure \ref{fig:signal_path} illustrates the data acquisition and signal processing chain used to recover the sawtooth-like spectral structure. The wideband noise signal generated by the noise source passes through a bandpass filter before being split into two branches by a splitter. Each branch connects to a separate data acquisition port via coaxial cables of varying lengths, replicating differences in spatial propagation lengths. This difference introduces a time delay of $30$ ns. After the analog signal is converted to digital format (ADC) by the two ports, it is channelized into a series of coarse channels. Given a specific time delay difference, the necessary phase shifting for each coarse channel is calculated and applied to the channelized signal of each branch. The phase-shifted signals from both branches are then summed and sent to the server for the second stage of channelization, where the spectrum is computed. Notably, the time delay used for calculating the phase shift differs by 0.3 ns from the delay resulting from the variations in cable lengths, simulating the offset of the source direction from the phase center. The resulting measured spectrum is depicted in Figure \ref{fig:actual_data}.

An evident sawtooth-like spectral structure is observed in the spectrum, mirroring the results from our actual observations of Cas A. It's worth noting that the large-scale variation is attributable to imperfections in the system's bandpass flatness, a factor not addressed in this study.

\section{The Cause of the Sawtooth-like Spectral Structure}
\label{sec:explaination}
\subsection{Theoretical Derivation of Station Spectral Response}
\label{ssec:deduction}
In this section, we will theoretically deduce the response of an antenna station to a monochromatic electromagnetic plane wave with frequency $\nu$ arriving from the direction $\mathbf{n}$.
The electric field intensity of a beam of plane waves incoming from the direction $\mathbf{n}$ with frequency $\nu$ can be expressed as a complex-valued function of time $t$ and spatial coordinates $\mathbf{x}$ as
\begin{equation}
	E(t,\mathbf{x})=\exp\left [2\pi i (\nu t + \frac{\nu\mathbf{n}\cdot \mathbf{x}}{c})\right ],
\end{equation}
\citep[e.g., ][]{born2013principles} where $c$ is the speed of light.
Note that $\mathbf{n}$ represents the direction of the radio source, which is opposite to the propagation direction or the wave vector $\mathbf{k}$.
To simplify the problem, we assume that all antennas are single-polarized, allowing the plane wave to be expressed as a scalar function.
The time-dependent term can be omitted so that
\begin{equation}
	E(\mathbf{x})=\exp\left (2\pi i \frac{\nu\mathbf{n}\cdot \mathbf{x}}{c}\right ).
\end{equation}
The coordinates of the $k$-th antenna in the station are denoted as $\mathbf{x}_k$, and the complex digital beamforming weight for each antenna is represented by $W_k(\nu)$.
Then, the voltage output of the station is calculated as
\begin{equation}
	\tilde{U}=\sum_k W_k(\nu) \exp \left (2\pi i \frac{\nu\mathbf{n}\cdot \mathbf{x}_k}{c}\right ).\label{eq:dbf}
\end{equation}
Note that we treat each individual antenna in the station as an omnidirectional antenna, which means that its beam pattern is independent of direction.

According to the theory of digital beamforming, when directing the station beam to a specific direction $\mathbf{n}_0$, the phase delays of the digital signals collected from all antennas induced by the plane wave incoming from this direction should be compensated to be equal \citep[e.g.,][]{rudnick1969digital, maranda1989efficient,2010arXiv1008.4047F}, so that $W_k$ can be calculated as
\begin{equation}
	W_k(\nu)=\exp\left [-2\pi i\frac{\nu^\prime(\nu)\mathbf{n}_0\cdot \mathbf{x}}{c}\right ],\label{eq:w}
\end{equation}
where $\nu^\prime$ is the frequency used to calculate the weights.
In an ideal condition, $\nu^\prime\equiv\nu$, and when the two-stage channelization is utilized, $\nu^\prime$ represents the central frequency of the coarse channel that encompasses frequency $\nu$ as
\begin{equation}
	\nu^\prime=\lfloor\frac{\nu}{B}+\frac{1}{2}\rfloor\times B,
\end{equation}
where $B$ is the bandwidth of a coarse channel.
By substituting equation \ref{eq:w} into equation \ref{eq:dbf}, we obtain
\begin{equation}
	\tilde{U}(\nu,\mathbf{n};\nu^\prime,\mathbf{n}_0)=\sum_k \exp\left [2\pi i\frac{(\nu \mathbf{n}-\nu^\prime \mathbf{n}_0)\cdot \mathbf{x}_k}{c}\right ].\label{eq:station_output}
\end{equation}

We can define an equivalent gain to represent the spectral response as
\begin{equation}
	g(\nu, \mathbf{n};\nu^\prime,\mathbf{n}_0)=\frac{|\tilde{U}(\nu, \mathbf{n};\nu^\prime,\mathbf{n}_0)|^2}{|\tilde{U}(\nu, \mathbf{n}_0;\nu,\mathbf{n}_0)|^2}\\
	=\left |\frac{1}{N}\sum_k \exp\left [2\pi i\frac{(\nu \mathbf{n}-\nu^\prime \mathbf{n}_0)\cdot \mathbf{x}_k}{c}\right ]\right |^2,\label{eq:equiv_gain}
\end{equation}
where $N$ is the number of antennas within a station.

\begin{figure}[hbt!]
	\centering
	\includegraphics[width=0.5\columnwidth]{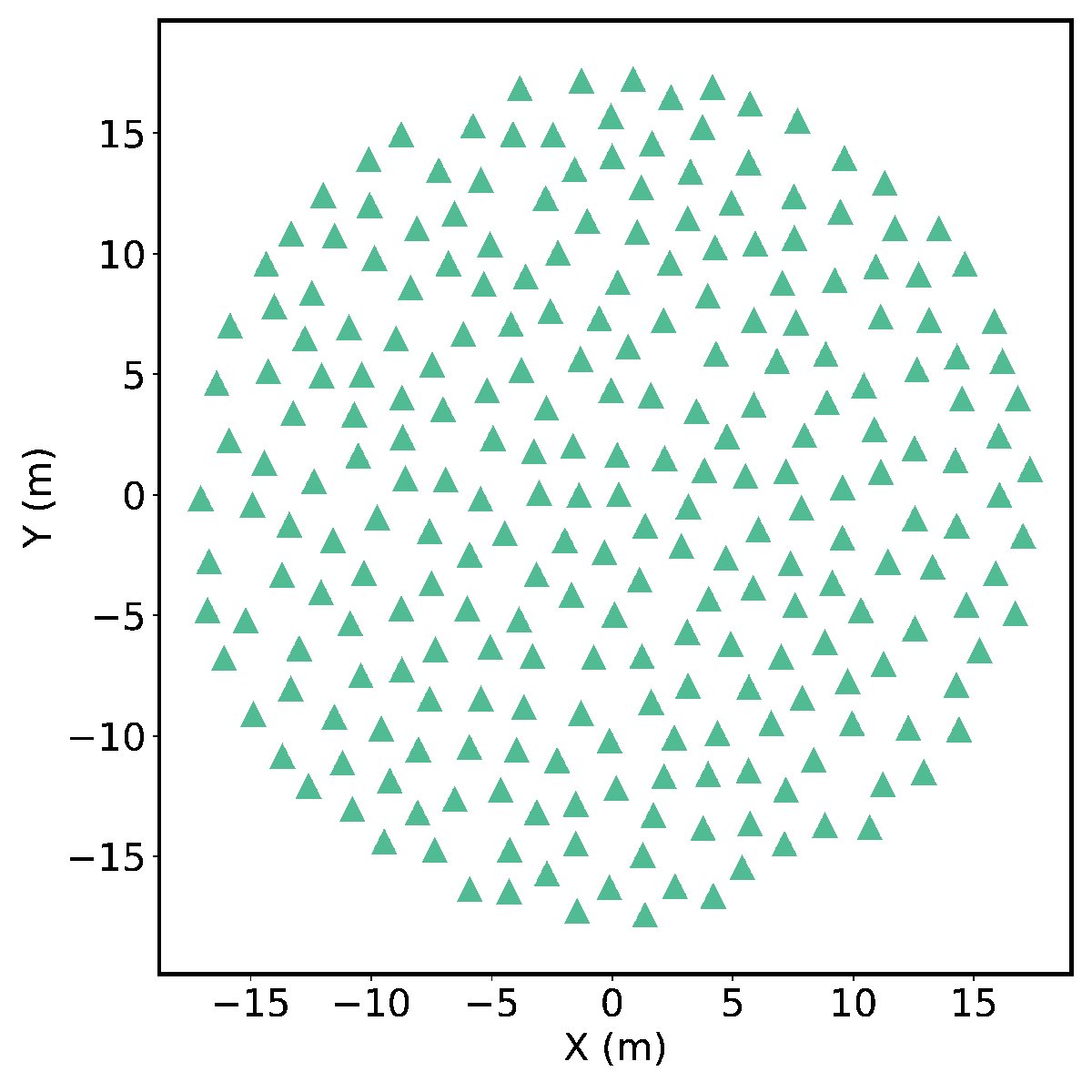}
	\caption{\label{fig:station} The station configuration that we used to calculate station response.}
\end{figure}

In order that the following analysis can be fit into the background of SKA-LFAA, we utilize some of SKA-LFAA's instrumental parameters, including the channelizer settings and the station configuration.
The coarse channel spacing is set to 781.25 kHz, as e.g., \cite{2017JAI.....641015C} and \cite{2020SPIE11450E..03C}. The frequency range is defined as 50-350 MHz (system requirement SYS\_REQ-2134, outlined in the "SKA PHASE 1 SYSTEM (LEVEL 1) REQUIREMENTS SPECIFICATION"\footnote{\url{https://www.skao.int/sites/default/files/documents/d3-SKA-TEL-SKO-0000008-Rev11_SKA1SystemRequirementSpecification.pdf}}).
We assume the station configuration satisfies system requirement SYS\_REQ-2140, i.e., a random configuration of 256 antennas distributed within a circular region of 35 m diameter, so that following analysis can be fit into the background of SKA-LFAA.
This station configuration is depicted in Figure \ref{fig:station}.
We define the $x$-axis as pointing eastward, the $y$-axis as pointing northward, and the $z$-axis as pointing towards the zenith.
By substituting the system parameters of SKA-LFAA into Equation \ref{eq:equiv_gain}, we further predict the station spectral response.

\begin{figure}[hbt!]
	\centering
	\includegraphics[width=\columnwidth]{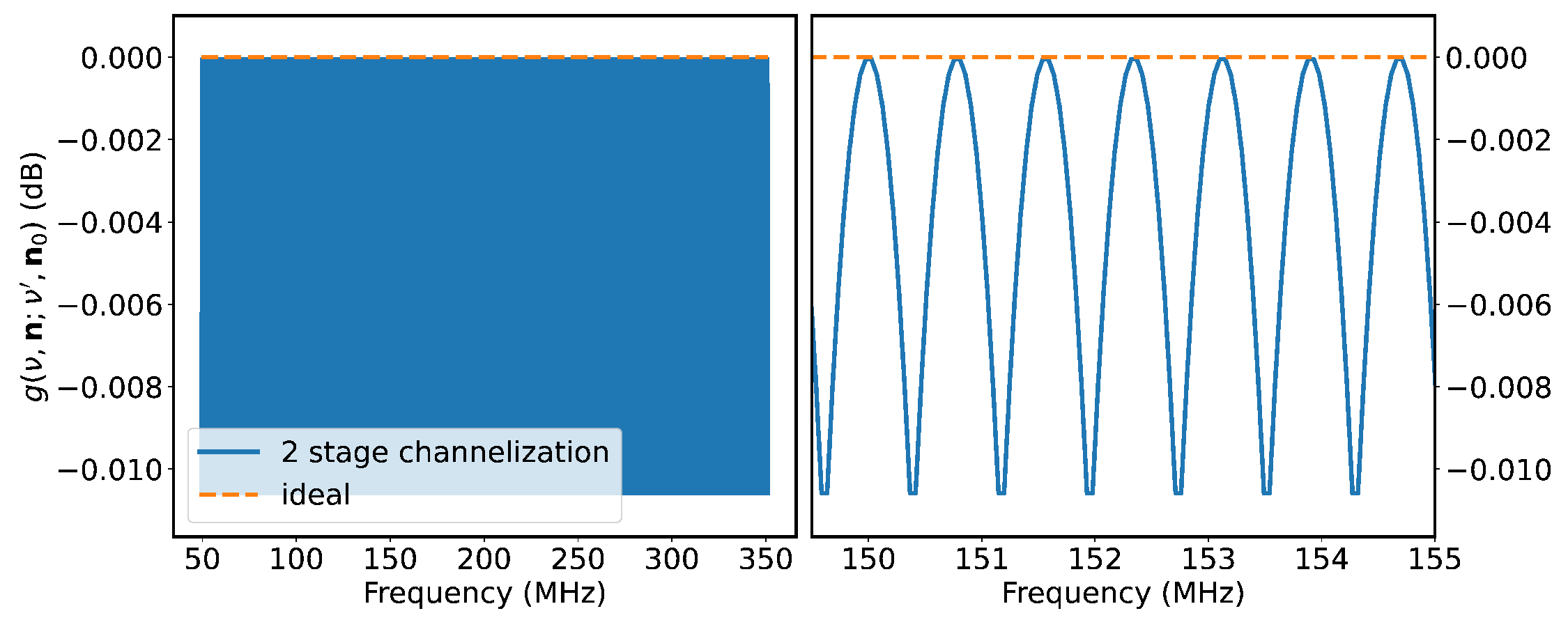}
	\caption{\label{fig:spec_center} The spectral response to a source positioned at the center of the beam. Left: covering the entire frequency range from 50 to 350 MHz; right: a close-up view providing detailed insights.}
\end{figure}

We set the desired beam pointing direction $\mathbf{n}_0$ to be  $[0, \frac{\sqrt{2}}{2}, \frac{\sqrt{2}}{2}]^\top$, which corresponds to an $\mathrm{Az}$ of $0^\circ$ and an $\mathrm{El}$ of $45^\circ$.
Initially, we investigate the spectral response to a point source located precisely at the center of the beam, i.e., $\mathbf{n}\equiv\mathbf{n}_0$.
Utilizing Equation \ref{eq:station_output}, we compute the spectral response, as illustrated in Figure \ref{fig:spec_center}.
The difference in equivalent gain within a coarse channel is approximately $0.012$ dB, corresponding to a relative measurement error of $2.8\times 10^{-3}$ before calibration.

\begin{figure}[hbt!]
	\centering
	\includegraphics[width=\columnwidth]{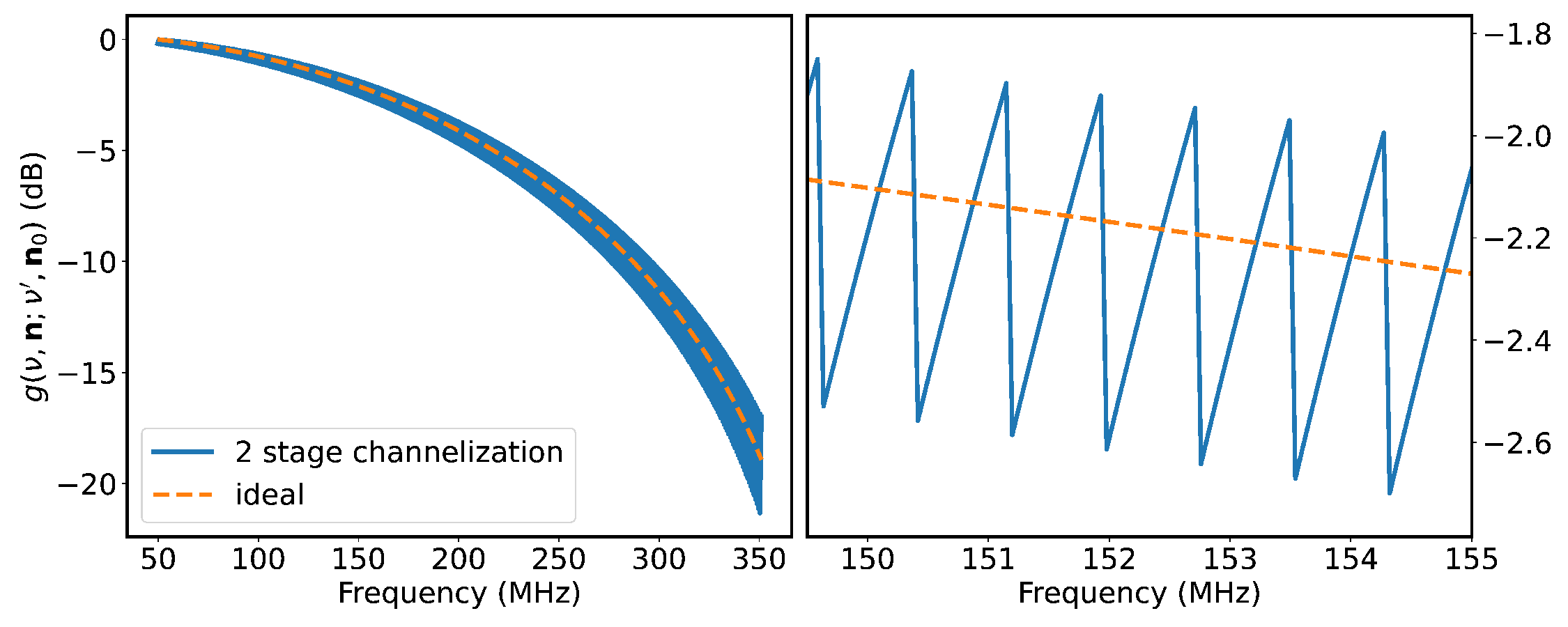}
	\caption{\label{fig:spec_off} The spectral response to a source positioned $2^\circ$ off the center of the beam, towards the zenith. Left: covering the entire frequency range from 50 to 350 MHz; right: a close-up view providing detailed insights.}
\end{figure}

We maintain the beam pointing direction at $\mathbf{n}_0$, with $\mathrm{Az}=0^\circ$ and $\mathrm{El}=45^\circ$, and then shift the hypothetical source away from the beam center to $\mathrm{Az}=0^\circ$ and $\mathrm{El}=47^\circ$, i.e., $2^\circ$ off-center towards the zenith.
The computed spectral response, as depicted in Figure \ref{fig:spec_off}, exhibits a clear sawtooth-like oscillating structure (SLOSS hereafter).
The difference in equivalent gain within a coarse channel near $150$ MHz is approximately $0.7$ dB, corresponding to a relative measurement error of $17\%$ before calibration.

\subsection{Full Numerical Simulation to LFAA DBF Subsystem}
\label{ssec:num_sim}
\begin{figure}[hbt!]
	\centering
	\includegraphics[width=\columnwidth]{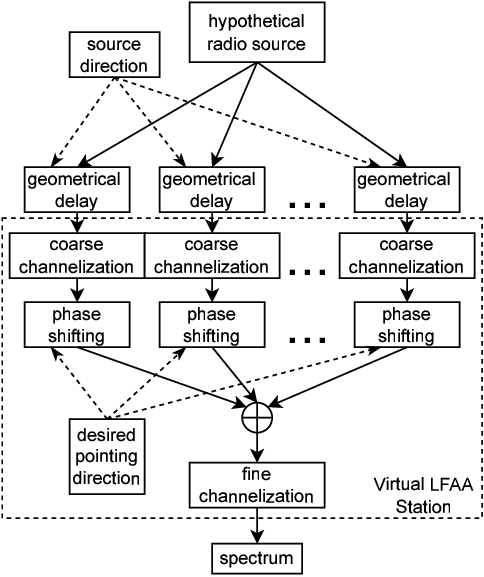}
	\caption{\label{fig:sim_cfg} The configuration of the simulated virtual LFAA station.}
\end{figure}

\begin{figure}[hbt!]
	\centering
	\includegraphics[width=\columnwidth]{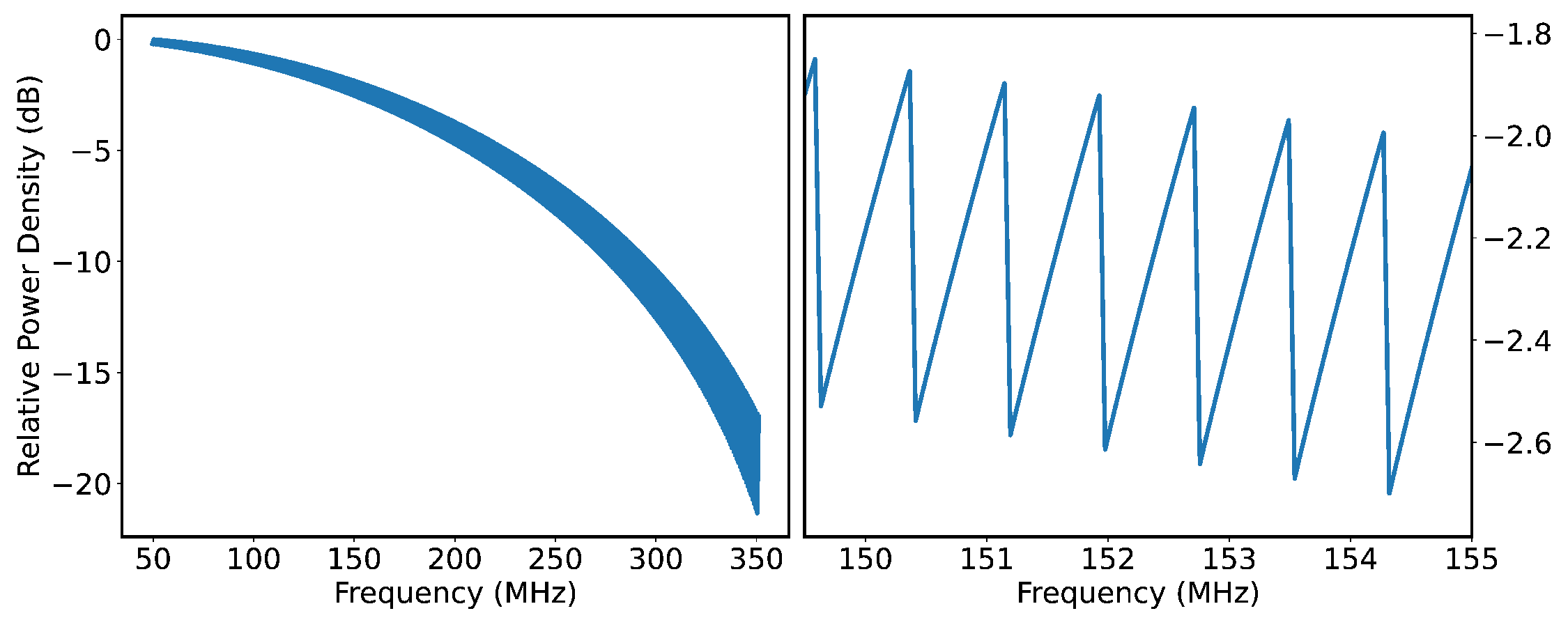}
	\caption{\label{fig:sim_auto} The output of a station to a white-noise point source that is placed $2^{\circ}$ off the beam center. The output is expressed in relative power density. The elevation angle of the station beam is set to be $45^\circ$. Left: covering the entire frequency range from 50 to 350 MHz; right: a close-up view providing detailed insights.}
\end{figure}

We further investigate the SLOSS effects through full numerical simulation.
For this purpose, a single hypothetical point radio source with a flat spectrum is utilized as the input.
The geometric delay difference between the individual antennas in the LFAA station is taken into account and emulated using a set of time-domain fractional delay filters.
The signal processing path of the simulation is illustrated in Figure \ref{fig:sim_cfg}.
We employ a two-stage channelization algorithm based on the polyphase filter bank (PFB) \citep[e.g.,][]{harris2003digital} algorithm.
In the first stage, the received radio signal is divided into a set of coarse channels with a bandwidth of 1562.5 kHz using an oversampling PFB.
The channel spacing of the PFB is set to 781.25 kHz, resulting in an oversampling ratio of 2.
Each coarse channel is further channelized using a critical sampling PFB, with fine channels outside the passband being discarded in the second stage.
This yields a set of fine channels with a bandwidth and channel spacing of 48.82 kHz.
We choose this fine channel bandwidth because it provides sufficient resolution to study the topic covered in this paper while maintaining an acceptable computational cost, although higher spectral resolution is feasible.
The station beam is directed towards $\mathrm{Az}=0^\circ$ and $\mathrm{El}=45^\circ$, while the hypothetical point radio source is positioned at $\mathrm{Az}=0^\circ$ and $\mathrm{El}=47^\circ$.
The spectrum computed from the station output is depicted in Figure \ref{fig:sim_auto}.

By focusing on a narrow band of the spectrum, the sawtooth-like oscillating spectral structure becomes clearly visible, as expected.
The amplitude of this oscillation feature is approximately 0.6 dB near 150 MHz.
Interestingly, the "period" of the oscillating feature precisely matches the coarse channel spacing of 781.25 kHz.
These findings are consistent with the predictions made in Section \ref{ssec:deduction}, as confirmed by the full numerical simulation.

\section{Discussion}
\label{sec:discussion}
In the preceding sections, we delineate the impact of the two-stage channelization strategy on the spectral response.
We observe that the amplitude responses of different fine channels within a single coarse channel are not uniform;
a systematic difference is discernible among these fine channels.
While this effect is relatively minor for a source positioned precisely at the beam center,
it becomes more pronounced for sources located off-center, posing greater challenges for mitigation.

\subsection{Influence on Visibility Phase}
When operating in an interferometric mode, the precision of phase measurements in visibility data becomes another critical consideration. Let's consider a scenario where we have two stations with identical configurations, forming a baseline $\mathbf{b}$. Referring to the notations used in Equation \ref{eq:station_output}, let $\mathbf{x}_k$ denote the coordinates of the $k$-th antenna in one station. Consequently, the coordinates of the $k$-th antenna in the other station are represented as $\mathbf{x}_k+\mathbf{b}$.

A hypothetical point source is positioned in the direction $\mathbf{n}$. The cross-correlation of the outputs of the two stations induced by this source is computed as
\begin{equation}
	\left <\tilde{U}_2\tilde{U}^*_1\right >=\left <\sum_k W_k(\nu) \exp\left [2\pi i \frac{\nu\mathbf{n}\cdot (\mathbf{x}_k+\mathbf{b})}{c}\right ]\right .\times \left . \left (\sum_k W_k(\nu) \exp\left [ 2\pi i \frac{\nu\mathbf{n}\cdot \mathbf{x}_k}{c} \right ]\right )^* \right >
\end{equation}
\begin{equation}
	=\left <\exp\left (2\pi i \frac{\nu\mathbf{n}\cdot \mathbf{b}}{c}\right )\left |\sum_k W_k(\nu) \exp\left [2\pi i \frac{\nu\mathbf{n}\cdot \mathbf{x}_k}{c}\right ]\right |^2 \right  >.\label{eq:corr}
\end{equation}
Although the amplitude of the cross-correlation i.e.,
\begin{equation}
	\left |\sum_k W_k(\nu) \exp\left [2\pi i \frac{\nu\mathbf{n}\cdot \mathbf{x}_k}{c}\right ]\right |^2
\end{equation}
is still influenced by the two-stage channelization mechanism, its phase i.e., $2\pi\nu\mathbf{n}\cdot \mathbf{b}/c$ is not influenced.

\subsection{Can the SLOSS Effects be Corrected by Primary Beam Calibration?}
The station's primary beam pattern can be predicted using Equation \ref{eq:equiv_gain}, thus theoretically, the effects induced by the two-stage channelization can be rectified through primary beam calibration.
However, in a real-world instrument, phase discrepancies arise from unknown errors in cable lengths, resulting in inaccuracies in the prediction of the primary beam pattern.
In this section, we assess the feasibility of correcting the SLOSS amidst these cable length errors.

\begin{figure}[hbt!]
	\centering
	\includegraphics[width=0.95\columnwidth]{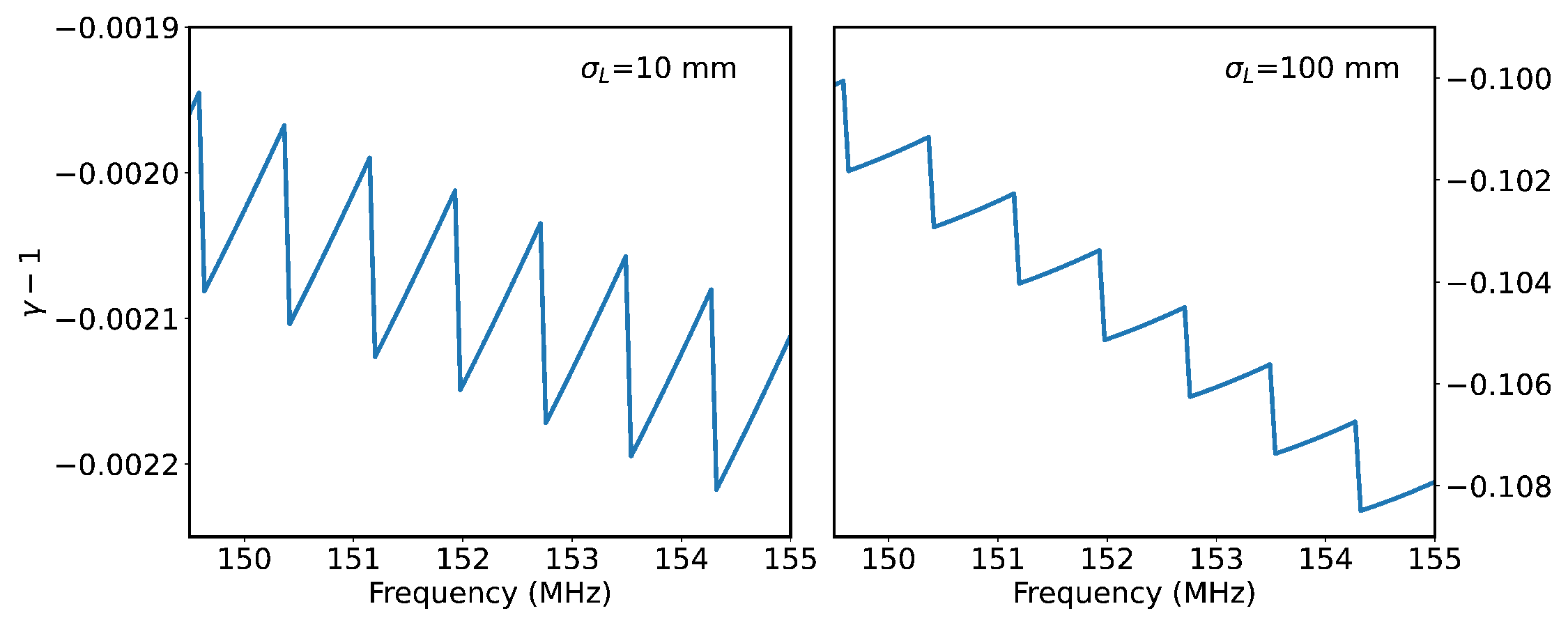}
	\caption{\label{fig:resid_err} Residual error after the primary beam calibration is performed.}
\end{figure}

We first simulate the station gain as
\begin{equation}
	g^\prime(\nu, \mathbf{n};\nu^\prime,\mathbf{n}_0, \mathbf{L})
	=\left |\frac{1}{N}\sum_k \exp\left [2\pi i\frac{\nu (\mathbf{n}\cdot \mathbf{x}_k-L_k)-\nu^\prime \mathbf{n}_0\cdot \mathbf{x}_k}{c}\right ]\right |^2 ,\label{eq:station_beam_with_cable_length_error}
\end{equation}
where $L_k$ is the uncorrected cable length error (the refraction index is regarded to be $1$) of the $k$-th antenna, and all the $L_k$'s are combined to form the vector $\mathbf{L}$.
We assume that the uncorrected cable length errors $L_k$ follow a zero-mean normal distribution, with a standard deviation of $\sigma_L$.
The corrected station gain, denoted as
\begin{equation}
	\gamma(\nu, \mathbf{n};\nu^\prime,\mathbf{n}_0, \mathbf{L})\equiv \frac{g^\prime(\nu, \mathbf{n};\nu^\prime,\mathbf{n}_0, \mathbf{L})}{g(\nu, \mathbf{n};\nu^\prime,\mathbf{n}_0)}
\end{equation}
serves as a metric for assessing the effectiveness of primary beam calibration. Ideally, if the station gain is perfectly corrected through primary beam calibration, $\gamma$ should equal $1$ for all spectral channels and directions. We illustrate the deviation of $\gamma$ from $1$ for all fine channels under conditions where $\sigma_L$ equals $10$ mm and $100$ mm in Figure \ref{fig:resid_err}. Evidently, for both conditions, the sawtooth-like spectral structure is evident.

In summary, although the effect can, in principle, be modeled using a theoretical station beam and mitigated through primary beam correction, the presence of cable delay errors prevents a perfect removal of the SLOSS-induced spectral structure in practice.

\subsection{Potential Influence on EoR Signal Detection}
Among the foreground subtraction methods for detecting the EoR signal, one approach relies on the assumption that the foreground is smooth and can be effectively approximated by low-degree polynomials \citep[e.g., ][]{2006ApJ...650..529W,2013ApJ...763...90W,2013ApJ...773...38G}. Similarly, in statistical measurement methods that do not necessitate foreground subtraction, the smoothness of the foreground spectra is assumed so that its significant scales can be distinguished from the fluctuating EoR neutral hydrogen 21 cm signals \citep[e.g., ][]{2016MNRAS.458.2928C,2014PhRvD..90b3018L}. Crucially, this assumption of intrinsic foreground smoothness implicitly requires a corresponding high degree of spectral smoothness in the instrumental bandpass \citep{2016MNRAS.461.3135B}. However, the sawtooth-shaped artificial spectral structure caused by the two-stage channelization strategy disrupts this assumption and could potentially contaminate the wave number regions that were presumed to be dominated by the EoR signals in statistical studies. Consequently, it is no longer viable to consider spectra across the coarse channel boundary as smooth. Nevertheless, it is still feasible to treat the spectra as smooth within one coarse channel. If the cable length of each antenna is precisely known, the primary beam correction can be employed to rectify the sawtooth-like structure. Unfortunately, if the cable lengths are not fully calibrated, these spectral structures will persist in the spectra of foreground sources, particularly those sources that are distant from the phase center. For a source located $2^\circ$ from the phase center, the amplitude of the sawtooth-like spectral structure is approximately $1\times10^{-4}$ and $2\times10^{-3}$ for $\sigma_L=10$ mm and $\sigma_L=100$ mm, respectively.
Given that the expected EoR signal amplitude is typically of the order of $10^{-4}$ relative to the foregrounds, and may be as low as $10^{-5}$ in some frequency ranges, residual gain errors exceeding this level can lead to coupling between foreground emission and the SLOSS-induced spectral structure. This may produce spurious spectral features whose amplitude becomes comparable to, or larger than, the EoR signal, thereby significantly complicating its detection. The precision requirements of EoR observations also set the calibration cadence: if the stability of cable delays satisfies these requirements, only limited calibration is required; otherwise, calibration must be performed for each observation.

In traditional radio interferometric imaging, direction-independent (DI) and direction-dependent (DD) calibration are routinely employed to solve for station-based complex gains. However, these techniques are primarily designed to calibrate the relative delays and phases between stations (on a per-baseline basis) and cannot directly resolve the individual delay errors of antennas within a single station. While it is theoretically possible to calibrate an intra-station antenna by isolating it for correlation against a reference station, such a procedure is impractical during standard scientific observations and would significantly compromise system sensitivity. Consequently, these unresolved intra-station cable-length errors remain embedded in the beamformed output, manifesting as the SLOSS effect described above.
If the station beam can be measured in practice, either using a drone-based measurement system or through astronomical observations combined with a sky model, the unknown antenna cable delays may potentially be calibrated.
In detail, the cable length errors $\mathbf{L}$ can be treated as unknown parameters to be determined in the following least-squares problem.
\begin{equation}
	\mathbf{L}=\arg \min_{\mathbf{L}} \sum_{i,j}\left | g^\prime(\nu_j, \mathbf{n}_i;\nu^\prime(\nu_j),\mathbf{n}_0, \mathbf{L}) - \tilde{g}(\nu_j, \mathbf{n}_i) \right |^2,
\end{equation}
where $\mathbf{n}_i$'s are a set of chosen directions in the field of view, $\nu_j$'s are a set of fine spectral channels, and $\tilde{g}$ is the station beam measured from actual data.
Given the estimated cable length errors $\mathbf{L}$, the station beam can be accurately predicted using Eq. \ref{eq:station_beam_with_cable_length_error} and subsequently used for primary beam calibration. However, for the operating frequency and station size of SKA-LFAA, it is difficult for drone-based measurements to satisfy the far-field condition. On the other hand, calibration using observational data combined with a sky model relies on prior assumptions about the sky model, and therefore cannot achieve fully blind calibration. A precise correction of the SLOSS effect is beyond the scope of this work and will be investigated in future studies.

\subsection{Reproducing Sawtooth-like Spectral Structure with \sc{Oskar}}
\label{sec:oskar}
The {\sc Oskar} code is utilized to conduct simulated observations for SKA. As of the end of 2023 (version 2.83, released on May 27, 2022), there is no officially documented built-in function in {\sc Oskar} to emulate the two-stage channelization mechanism in the SKA-LFAA beamformer. As demonstrated earlier, this mechanism can produce observable effects, necessitating an exploration of its influence on scientific data reduction before actual data becomes available. In this section, we present the method to replicate this effect in {\sc Oskar}.

First, we bypass {\sc Oskar}'s internal pointing calculation by utilizing a station pointing file that contains a single line as
\begin{verbatim}
    * AZEL  0.0 90.0
\end{verbatim}
representing the zenith, so that no automatically calculated phase delay is applied to the antennas.

\begin{figure*}
	\centering
	\includegraphics[width=0.8\textwidth]{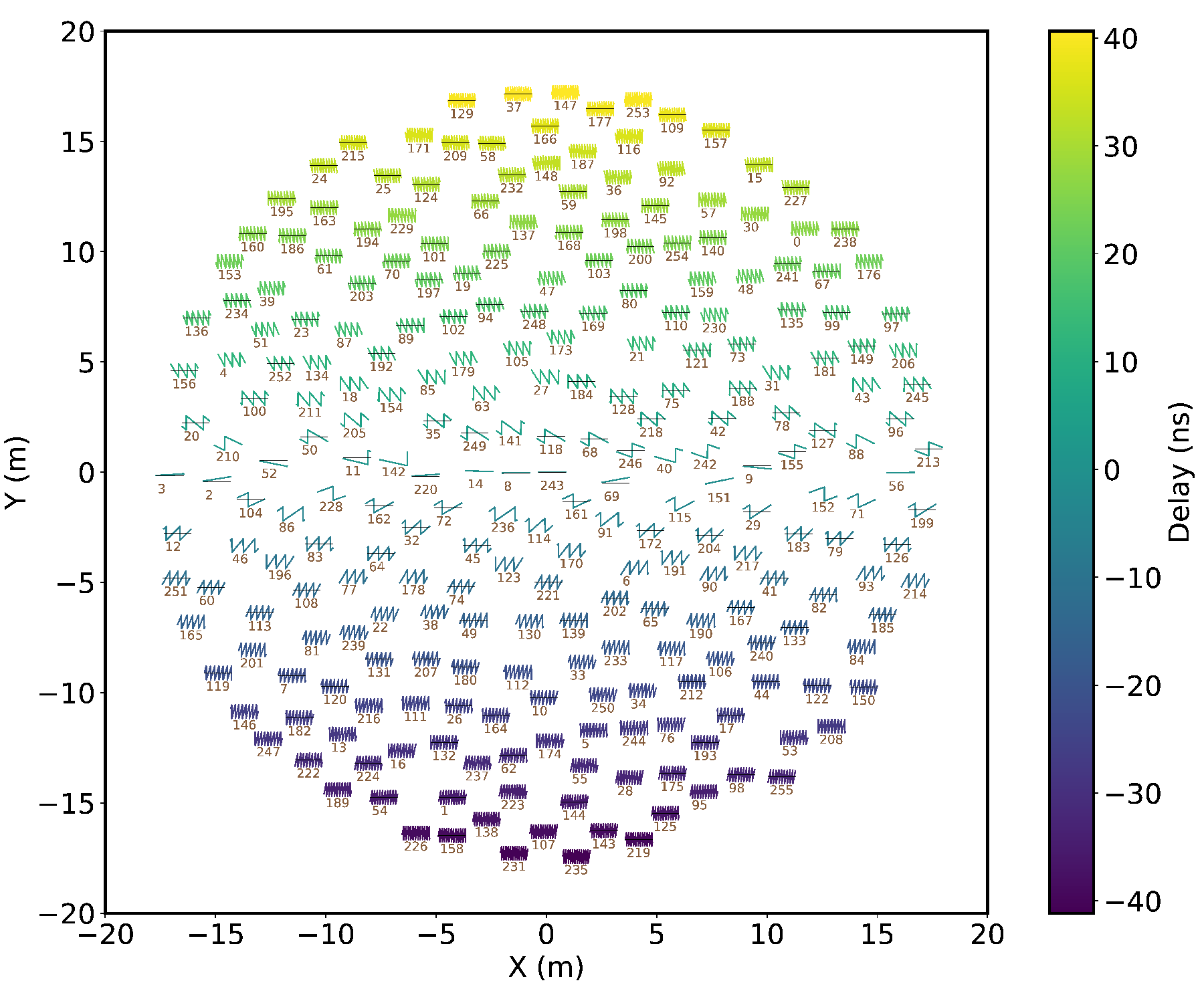}
	
	\caption{\label{fig:phase_vs_freq} The phase of the complex gain as a function of the spectral channel of each antenna in the station. The colors of the curves encode the required delays to form the desired beam.}
\end{figure*}

\begin{figure}[hbt!]
	\centering
	\includegraphics[width=0.95\columnwidth]{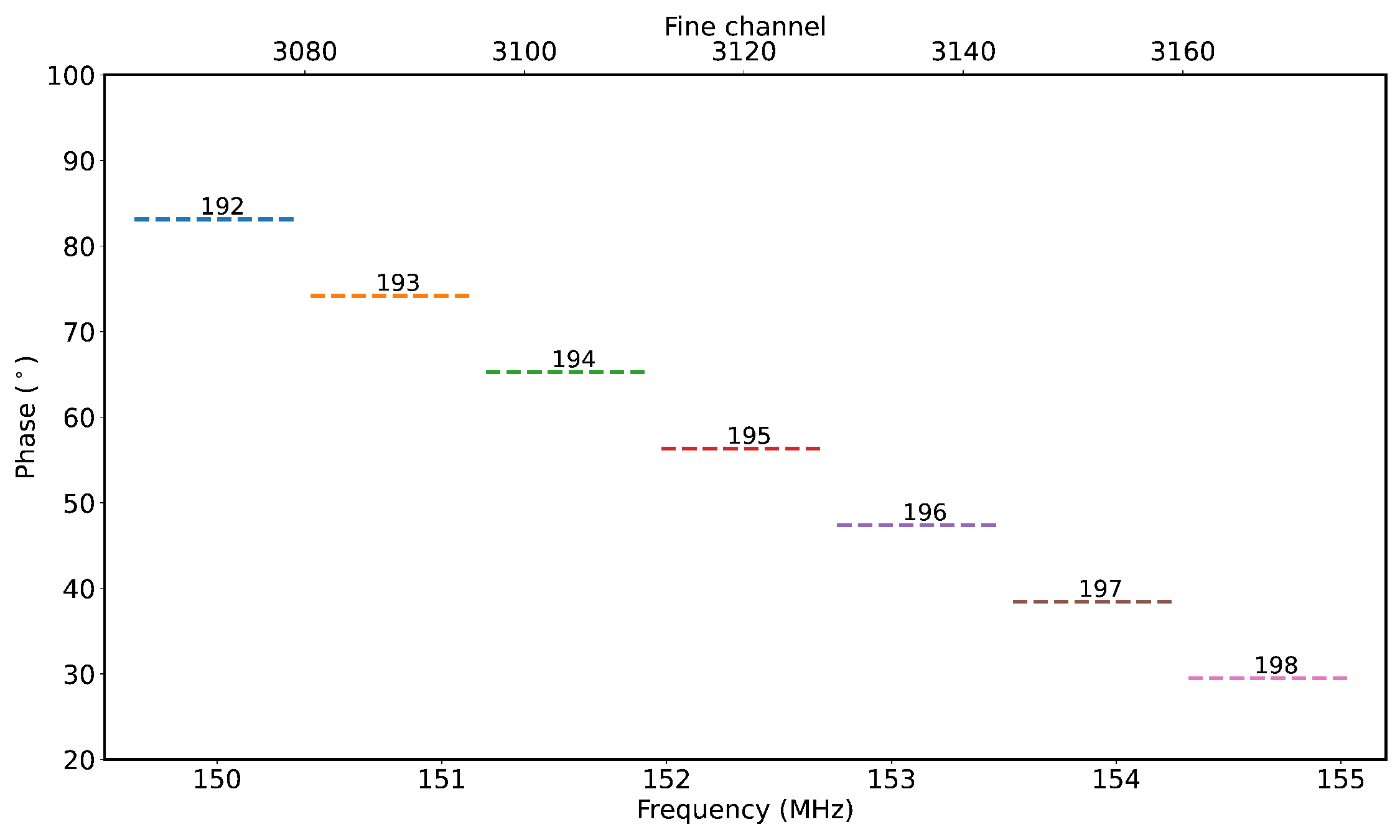}
	\caption{\label{fig:fig:phase_vs_freq_1_ant}The phase of the complex gain as a function of frequency of the antenna No. 232. The frequency range corresponding to the 192nd to 198th coarse channels is shown.}
\end{figure}

Next, we calculate the complex gain for each antenna and each fine spectral channel using Equation \ref{eq:w}, and then write it into a gain file, following the guidelines provided in the {\sc Oskar} documentation\footnote{\url{https://ska-telescope.gitlab.io/sim/oskar/telescope_model/telescope_model.html?highlight=gain_model\#telescope-gain-model}}.
The beam pointing direction remains set to be $\mathrm{Az}=0^\circ$, $\mathrm{El}=45^\circ$.
The relationship between the phase of the complex gain and the spectral channel of each antenna is depicted in Figure \ref{fig:phase_vs_freq}.
To provide a more detailed view, we plot the same relationship for antenna No. 232 within a narrower frequency range in Figure \ref{fig:fig:phase_vs_freq_1_ant}.

\begin{figure}[hbt!]
	\centering
	\includegraphics[width=0.5\columnwidth]{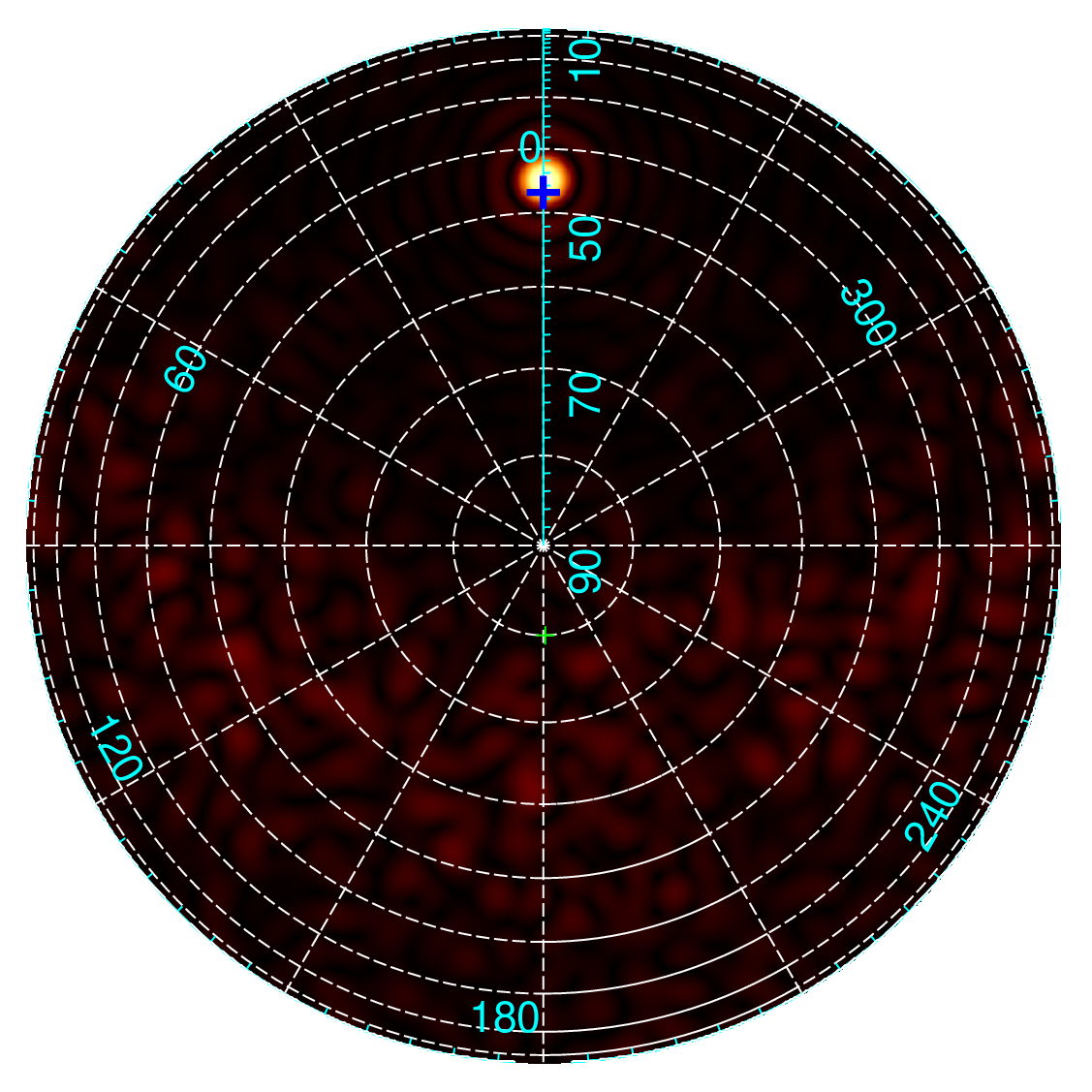}
	\caption{\label{fig:oskar_beam} The beam pattern of 150 MHz spectral channel simulated with the {\sc Oskar oskar\_sim\_beam\_pattern} tool. The blue cross denotes the direction where a hypothetical point source with a flat spectrum is placed to examine the spectral response}
\end{figure}

\begin{figure}[hbt!]
	\centering
	\includegraphics[width=\columnwidth]{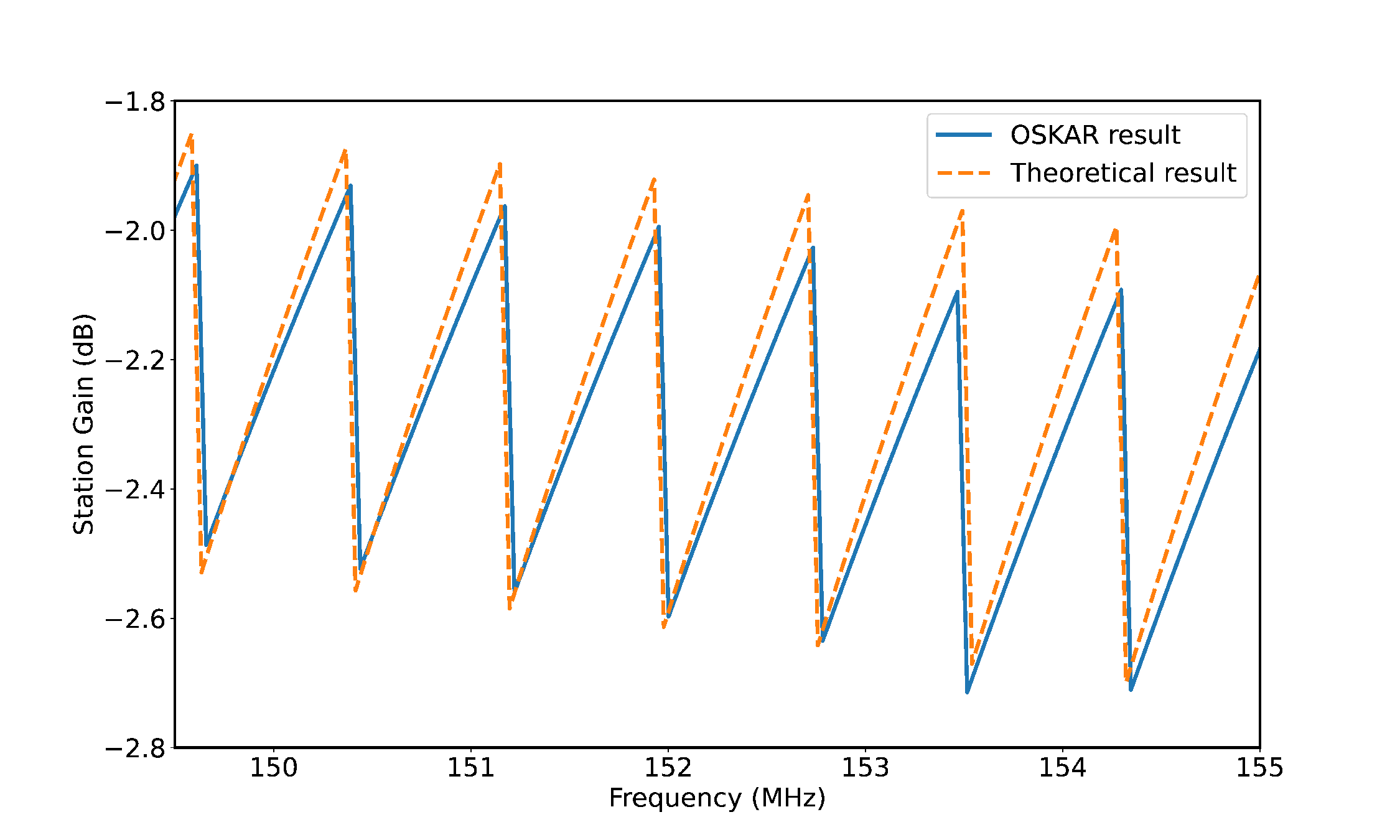}
	\caption{\label{fig:oskar_spectrum} Station beam gain at $\mathrm{Az}=0^\circ$, $\mathrm{El}=47^\circ$ for each fine channel, computed with {\sc Oskar} (blue solid line), compared with the theoretical prediction from Eq.~\ref{eq:equiv_gain} (orange dashed line).}
\end{figure}

We employ the {\sc oskar\_sim\_beam\_pattern} tool from the {\sc Oskar} package to compute the station beam. The simulated beam pattern at 150 MHz is depicted in Figure \ref{fig:oskar_beam}. To ensure consistency, the gains of all spectral channels at the beam center are renormalized to $1$ (i.e., 0 dB). We then extract the station beam gain for each fine channel in the direction $\mathrm{Az}=0^\circ$, $\mathrm{El}=47^\circ$, as illustrated in Fig.~\ref{fig:oskar_spectrum}. The SLOSS effect is clearly evident, closely resembling the results presented in Section~\ref{sec:explaination}. We also include the theoretical result calculated using Eq.~\ref{eq:equiv_gain} in the same figure. The slight mismatch between the {\sc Oskar} result and the theoretical prediction can be attributed to the fact that the sky map produced by {\sc Oskar} is discrete, so that the selected direction for extracting the spectrum is not exact.

From a practical simulation standpoint, the computational and storage requirements for implementing this custom gain approach are relatively modest. For a single station, the total file size for the complex gains of all antennas is approximately 25 MB. During an interferometric simulation, the complex gains must be updated at each integration time step to account for the changing pointing directions. While the memory footprint and I/O overhead for these gain files are negligible compared to the visibility calculation itself, the process requires repeated calls to external user-defined routines to compute the DBF weights based on Equation \ref{eq:w}. This frequent context switching may introduce significant computational overhead for large-scale, wide-field EoR simulations involving thousands of sources and long integration times. Therefore, while our current method effectively reproduces the SLOSS effect for validation purposes, integrating this functionality directly into the OSKAR core would be the most efficient path for researchers planning high-performance, full-array simulations.

\section{Conclusions}
\label{sec:conclusions}
In this work, we have presented the design and commissioning of the Ng21CMA telescope, which serves as a research platform for investigating the instrumental complexities of SKA-LFAA-like architectures. The implementation of a modernized digital backend supporting real-time digital beamforming has been validated through both interferometric imaging and pulsar observations, confirming the system's operational stability and sensitivity.

A central finding of our instrumental assessment is the impact of the two-stage channelization strategy on spectral structure. We demonstrate that this architecture, while computationally efficient, introduces a direction-dependent sawtooth-like spectral structure that challenges the spectral smoothness required for EoR detection. Our analysis of these artifacts and the exploration of potential calibration requirements provide a practical case study for future radio telescopes with similar digital designs, underscoring the need for meticulous instrumental characterization in next-generation aperture arrays.

\begin{acknowledgements}
	This work is supported by National SKA Program of China No. 2020SKA0110200 and 2020SKA0110100. 
	This work is supported by the National Natural Science Foundation of China (NSFC) under grant No. 12573097.
\end{acknowledgements}

\bibliographystyle{raa}
\bibliography{ms}

@article{2017JAI.....641015C,
    author        = {{Comoretto}, Gianni and {Chiello}, Riccardo and {Roberts}, Matt and {Halsall}, Rob and {Adami}, Kristian Zarb and {Alderighi}, Monica and {Aminaei}, Amin and {Baker}, Jeremy and {Belli}, Carolina and {Chiarucci}, Simone and {D'Angelo}, Sergio and {De Marco}, Andrea and {Mura}, Gabriele Dalle and {Magro}, Alessio and {Mattana}, Andrea and {Monari}, Jader and {Naldi}, Giovanni and {Pastore}, Sandro and {Perini}, Federico and {Poloni}, Marco and {Pupillo}, Giuseppe and {Rusticelli}, Simone and {Schiaffino}, Marco and {Schillir{\`o}}, Francesco and {Zaccaro}, Emanuele},
    title         = {{The Signal Processing Firmware for the Low Frequency Aperture Array}},
    journal       = {Journal of Astronomical Instrumentation},
    year          = 2017,
    month         = mar,
    volume        = {6},
    number        = {1},
    eid           = {1641015},
    pages         = {1641015},
    doi           = {10.1142/S2251171716410154},
    archiveprefix = {arXiv},
    eprint        = {2002.10307},
    primaryclass  = {astro-ph.IM},
    adsurl        = {https://ui.adsabs.harvard.edu/abs/2017JAI.....641015C}
}

@inproceedings{2020SPIE11450E..03C,
    author    = {{Chiarucci}, Simone and {Comoretto}, Giovanni},
    title     = {{An end-to-end model for the correlator and beamformer of the Square Kilometer Array Low Frequency Aperture Array}},
    booktitle = {Society of Photo-Optical Instrumentation Engineers (SPIE) Conference Series},
    year      = 2020,
    series    = {Society of Photo-Optical Instrumentation Engineers (SPIE) Conference Series},
    volume    = {11450},
    month     = dec,
    eid       = {1145003},
    pages     = {1145003},
    doi       = {10.1117/12.2560838},
    adsurl    = {https://ui.adsabs.harvard.edu/abs/2020SPIE11450E..03C}
}

@article{2017MmSAI..88..154C,
    author   = {{Comoretto}, G. and {Chiarucci}, S. and {Belli}, C.},
    title    = {{Radioastronomic signal processing cores for the SKA radio telescope}},
    journal  = {\memsai},
    year     = 2017,
    month    = jan,
    volume   = {88},
    pages    = {154},
    adsurl   = {https://ui.adsabs.harvard.edu/abs/2017MmSAI..88..154C}
}

@article{2016PASA...33...19T,
    author        = {{Trott}, Cathryn M. and {Wayth}, Randall B.},
    title         = {{Spectral Calibration Requirements of Radio Interferometers for Epoch of Reionisation Science with the SKA}},
    journal       = {\pasa},
    year          = 2016,
    month         = may,
    volume        = {33},
    eid           = {e019},
    pages         = {e019},
    doi           = {10.1017/pasa.2016.18},
    archiveprefix = {arXiv},
    eprint        = {1604.03273},
    primaryclass  = {astro-ph.IM},
    adsurl        = {https://ui.adsabs.harvard.edu/abs/2016PASA...33...19T}
}

@article{2006ApJ...650..529W,
    author        = {{Wang}, Xiaomin and {Tegmark}, Max and {Santos}, M{\'a}rio G. and {Knox}, Lloyd},
    title         = {{21 cm Tomography with Foregrounds}},
    journal       = {\apj},
    year          = 2006,
    month         = oct,
    volume        = {650},
    number        = {2},
    pages         = {529-537},
    doi           = {10.1086/506597},
    archiveprefix = {arXiv},
    eprint        = {astro-ph/0501081},
    primaryclass  = {astro-ph},
    adsurl        = {https://ui.adsabs.harvard.edu/abs/2006ApJ...650..529W}
}

@article{2006PhR...433..181F,
    author        = {{Furlanetto}, Steven R. and {Oh}, S. Peng and {Briggs}, Frank H.},
    title         = {{Cosmology at low frequencies: The 21 cm transition and the high-redshift Universe}},
    journal       = {\physrep},
    year          = 2006,
    month         = oct,
    volume        = {433},
    number        = {4-6},
    pages         = {181-301},
    doi           = {10.1016/j.physrep.2006.08.002},
    archiveprefix = {arXiv},
    eprint        = {astro-ph/0608032},
    primaryclass  = {astro-ph},
    adsurl        = {https://ui.adsabs.harvard.edu/abs/2006PhR...433..181F}
}

@phdthesis{2015PhDT.......574S,
    author  = {{Sinclair}, David Robert},
    title   = {{A Study of the Square Kilometre Array Low-Frequency Aperture Array}},
    school  = {University of Oxford},
    year    = 2015,
    month   = jan,
    adsurl  = {https://ui.adsabs.harvard.edu/abs/2015PhDT.......574S}
}

@article{2010arXiv1008.4051K,
    author        = {{Khlebnikov}, Vasily A. and {Zarb-Adami}, Kristian and {Armstrong}, Richard P. and {Jones}, Michael E.},
    title         = {{All-Digital Wideband Space-Frequency Beamforming for the SKA Aperture Array}},
    journal       = {arXiv e-prints},
    year          = 2010,
    month         = aug,
    eid           = {arXiv:1008.4051},
    pages         = {arXiv:1008.4051},
    archiveprefix = {arXiv},
    eprint        = {1008.4051},
    primaryclass  = {astro-ph.IM},
    adsurl        = {https://ui.adsabs.harvard.edu/abs/2010arXiv1008.4051K}
}

@article{harris2003digital,
    title     = {Digital receivers and transmitters using polyphase filter banks for wireless communications},
    author    = {Harris, Fredric J and Dick, Chris and Rice, Michael},
    journal   = {IEEE transactions on microwave theory and techniques},
    volume    = {51},
    number    = {4},
    pages     = {1395--1412},
    year      = {2003},
    publisher = {IEEE}
}

@book{born2013principles,
    title     = {Principles of optics: electromagnetic theory of propagation, interference and diffraction of light},
    author    = {Born, Max and Wolf, Emil},
    year      = {2013},
    publisher = {Elsevier}
}

@article{rudnick1969digital,
    title     = {Digital beamforming in the frequency domain},
    author    = {Rudnick, Philip},
    journal   = {The Journal of the Acoustical Society of America},
    volume    = {46},
    number    = {5A},
    pages     = {1089--1090},
    year      = {1969},
    publisher = {Acoustical Society of America}
}

@article{maranda1989efficient,
    title     = {Efficient digital beamforming in the frequency domain},
    author    = {Maranda, Brian},
    journal   = {The Journal of the Acoustical Society of America},
    volume    = {86},
    number    = {5},
    pages     = {1813--1819},
    year      = {1989},
    publisher = {Acoustical Society of America}
}

@article{2010arXiv1008.4047F,
    author        = {{Faulkner}, A J and {Zarb Adami}, K and {Bij de Vaate}, J.~G. and {Kant}, G.~W. and {Pickard}, P.},
    title         = {{Beamforming Techniques for Large-N Aperture Arrays}},
    journal       = {arXiv e-prints},
    year          = 2010,
    month         = aug,
    eid           = {arXiv:1008.4047},
    pages         = {arXiv:1008.4047},
    archiveprefix = {arXiv},
    eprint        = {1008.4047},
    primaryclass  = {astro-ph.IM},
    adsurl        = {https://ui.adsabs.harvard.edu/abs/2010arXiv1008.4047F}
}

@article{2013ApJ...763...90W,
    author        = {{Wang}, Jingying and {Xu}, Haiguang and {An}, Tao and {Gu}, Junhua and {Guo}, Xueying and {Li}, Weitian and {Wang}, Yu and {Liu}, Chengze and {Martineau-Huynh}, Olivier and {Wu}, Xiang-Ping},
    title         = {{Exploring the Cosmic Reionization Epoch in Frequency Space: An Improved Approach to Remove the Foreground in 21 cm Tomography}},
    journal       = {\apj},
    year          = 2013,
    month         = feb,
    volume        = {763},
    number        = {2},
    eid           = {90},
    pages         = {90},
    doi           = {10.1088/0004-637X/763/2/90},
    archiveprefix = {arXiv},
    eprint        = {1211.6450},
    primaryclass  = {astro-ph.CO},
    adsurl        = {https://ui.adsabs.harvard.edu/abs/2013ApJ...763...90W}
}

@article{2013ApJ...773...38G,
    author        = {{Gu}, Junhua and {Xu}, Haiguang and {Wang}, Jingying and {An}, Tao and {Chen}, Wen},
    title         = {{The Application of Continuous Wavelet Transform Based Foreground Subtraction Method in 21 cm Sky Surveys}},
    journal       = {\apj},
    year          = 2013,
    month         = aug,
    volume        = {773},
    number        = {1},
    eid           = {38},
    pages         = {38},
    doi           = {10.1088/0004-637X/773/1/38},
    archiveprefix = {arXiv},
    eprint        = {1306.5378},
    primaryclass  = {astro-ph.CO},
    adsurl        = {https://ui.adsabs.harvard.edu/abs/2013ApJ...773...38G}
}

@article{2016MNRAS.458.2928C,
    author   = {{Chapman}, Emma and {Zaroubi}, Saleem and {Abdalla}, Filipe B. and {Dulwich}, Fred and {Jeli{\'c}}, Vibor and {Mort}, Benjamin},
    title    = {{The effect of foreground mitigation strategy on EoR window recovery}},
    journal  = {\mnras},
    year     = 2016,
    month    = may,
    volume   = {458},
    number   = {3},
    pages    = {2928-2939},
    doi      = {10.1093/mnras/stw161},
    adsurl   = {https://ui.adsabs.harvard.edu/abs/2016MNRAS.458.2928C}
}

@article{2014PhRvD..90b3018L,
    author        = {{Liu}, Adrian and {Parsons}, Aaron R. and {Trott}, Cathryn M.},
    title         = {{Epoch of reionization window. I. Mathematical formalism}},
    journal       = {\prd},
    year          = 2014,
    month         = jul,
    volume        = {90},
    number        = {2},
    eid           = {023018},
    pages         = {023018},
    doi           = {10.1103/PhysRevD.90.023018},
    archiveprefix = {arXiv},
    eprint        = {1404.2596},
    primaryclass  = {astro-ph.CO},
    adsurl        = {https://ui.adsabs.harvard.edu/abs/2014PhRvD..90b3018L}
}

@inproceedings{5340196,
    author    = {Moreira, Pedro and Serrano, Javier and Wlostowski, Tomasz and Loschmidt, Patrick and Gaderer, Georg},
    booktitle = {2009 International Symposium on Precision Clock Synchronization for Measurement, Control and Communication},
    title     = {White rabbit: Sub-nanosecond timing distribution over ethernet},
    year      = {2009},
    volume    = {},
    number    = {},
    pages     = {1-5},
    doi       = {10.1109/ISPCS.2009.5340196}
}

@ARTICLE{2016MNRAS.461.3135B,
       author = {{Barry}, N. and {Hazelton}, B. and {Sullivan}, I. and {Morales}, M.~F. and {Pober}, J.~C.},
        title = "{Calibration requirements for detecting the 21 cm epoch of reionization power spectrum and implications for the SKA}",
      journal = {\mnras},
         year = 2016,
        month = sep,
       volume = {461},
       number = {3},
        pages = {3135-3144},
          doi = {10.1093/mnras/stw1380},
archivePrefix = {arXiv},
       eprint = {1603.00607},
 primaryClass = {astro-ph.IM},
       adsurl = {https://ui.adsabs.harvard.edu/abs/2016MNRAS.461.3135B}
}

@ARTICLE{2015ExA....39...73P,
       author = {{Prabu}, Thiagaraj and {Srivani}, K.~S. and {Roshi}, D. Anish and {Kamini}, P.~A. and {Madhavi}, S. and {Emrich}, David and {Crosse}, Brian and {Williams}, Andrew J. and {Waterson}, Mark and {Deshpande}, Avinash A. and {Shankar}, N. Udaya and {Subrahmanyan}, Ravi and {Briggs}, Frank H. and {Goeke}, Robert F. and {Tingay}, Steven J. and {Johnston-Hollitt}, Melanie and {R}, Gopalakrishna M. and {Morgan}, Edward H. and {Pathikulangara}, Joseph and {Bunton}, John D. and {Hampson}, Grant and {Williams}, Christopher and {Ord}, Stephen M. and {Wayth}, Randall B. and {Kumar}, Deepak and {Morales}, Miguel F. and {deSouza}, Ludi and {Kratzenberg}, Eric and {Pallot}, D. and {McWhirter}, Russell and {Hazelton}, Bryna J. and {Arcus}, Wayne and {Barnes}, David G. and {Bernardi}, Gianni and {Booler}, T. and {Bowman}, Judd D. and {Cappallo}, Roger J. and {Corey}, Brian E. and {Greenhill}, Lincoln J. and {Herne}, David and {Hewitt}, Jacqueline N. and {Kaplan}, David L. and {Kasper}, Justin C. and {Kincaid}, Barton B. and {Koenig}, Ronald and {Lonsdale}, Colin J. and {Lynch}, Mervyn J. and {Mitchell}, Daniel A. and {Oberoi}, Divya and {Remillard}, Ronald A. and {Rogers}, Alan E. and {Salah}, Joseph E. and {Sault}, Robert J. and {Stevens}, Jamie B. and {Tremblay}, S. and {Webster}, Rachel L. and {Whitney}, Alan R. and {Wyithe}, Stuart B.},
        title = "{A digital-receiver for the MurchisonWidefield Array}",
      journal = {Experimental Astronomy},
         year = 2015,
        month = mar,
       volume = {39},
       number = {1},
        pages = {73-93},
          doi = {10.1007/s10686-015-9444-3},
archivePrefix = {arXiv},
       eprint = {1502.05015},
 primaryClass = {astro-ph.IM},
       adsurl = {https://ui.adsabs.harvard.edu/abs/2015ExA....39...73P}
}

@ARTICLE{2013A&A...556A...2V,
       author = {{van Haarlem}, M.~P. and {Wise}, M.~W. and {Gunst}, A.~W. and {Heald}, G. and {McKean}, J.~P. and {Hessels}, J.~W.~T. and {de Bruyn}, A.~G. and {Nijboer}, R. and {Swinbank}, J. and {Fallows}, R. and {Brentjens}, M. and {Nelles}, A. and {Beck}, R. and {Falcke}, H. and {Fender}, R. and {H{\"o}randel}, J. and {Koopmans}, L.~V.~E. and {Mann}, G. and {Miley}, G. and {R{\"o}ttgering}, H. and {Stappers}, B.~W. and {Wijers}, R.~A.~M.~J. and {Zaroubi}, S. and {van den Akker}, M. and {Alexov}, A. and {Anderson}, J. and {Anderson}, K. and {van Ardenne}, A. and {Arts}, M. and {Asgekar}, A. and {Avruch}, I.~M. and {Batejat}, F. and {B{\"a}hren}, L. and {Bell}, M.~E. and {Bell}, M.~R. and {van Bemmel}, I. and {Bennema}, P. and {Bentum}, M.~J. and {Bernardi}, G. and {Best}, P. and {B{\^\i}rzan}, L. and {Bonafede}, A. and {Boonstra}, A.-J. and {Braun}, R. and {Bregman}, J. and {Breitling}, F. and {van de Brink}, R.~H. and {Broderick}, J. and {Broekema}, P.~C. and {Brouw}, W.~N. and {Br{\"u}ggen}, M. and {Butcher}, H.~R. and {van Cappellen}, W. and {Ciardi}, B. and {Coenen}, T. and {Conway}, J. and {Coolen}, A. and {Corstanje}, A. and {Damstra}, S. and {Davies}, O. and {Deller}, A.~T. and {Dettmar}, R.-J. and {van Diepen}, G. and {Dijkstra}, K. and {Donker}, P. and {Doorduin}, A. and {Dromer}, J. and {Drost}, M. and {van Duin}, A. and {Eisl{\"o}ffel}, J. and {van Enst}, J. and {Ferrari}, C. and {Frieswijk}, W. and {Gankema}, H. and {Garrett}, M.~A. and {de Gasperin}, F. and {Gerbers}, M. and {de Geus}, E. and {Grie{\ss}meier}, J.-M. and {Grit}, T. and {Gruppen}, P. and {Hamaker}, J.~P. and {Hassall}, T. and {Hoeft}, M. and {Holties}, H.~A. and {Horneffer}, A. and {van der Horst}, A. and {van Houwelingen}, A. and {Huijgen}, A. and {Iacobelli}, M. and {Intema}, H. and {Jackson}, N. and {Jelic}, V. and {de Jong}, A. and {Juette}, E. and {Kant}, D. and {Karastergiou}, A. and {Koers}, A. and {Kollen}, H. and {Kondratiev}, V.~I. and {Kooistra}, E. and {Koopman}, Y. and {Koster}, A. and {Kuniyoshi}, M. and {Kramer}, M. and {Kuper}, G. and {Lambropoulos}, P. and {Law}, C. and {van Leeuwen}, J. and {Lemaitre}, J. and {Loose}, M. and {Maat}, P. and {Macario}, G. and {Markoff}, S. and {Masters}, J. and {McFadden}, R.~A. and {McKay-Bukowski}, D. and {Meijering}, H. and {Meulman}, H. and {Mevius}, M. and {Middelberg}, E. and {Millenaar}, R. and {Miller-Jones}, J.~C.~A. and {Mohan}, R.~N. and {Mol}, J.~D. and {Morawietz}, J. and {Morganti}, R. and {Mulcahy}, D.~D. and {Mulder}, E. and {Munk}, H. and {Nieuwenhuis}, L. and {van Nieuwpoort}, R. and {Noordam}, J.~E. and {Norden}, M. and {Noutsos}, A. and {Offringa}, A.~R. and {Olofsson}, H. and {Omar}, A. and {Orr{\'u}}, E. and {Overeem}, R. and {Paas}, H. and {Pandey-Pommier}, M. and {Pandey}, V.~N. and {Pizzo}, R. and {Polatidis}, A. and {Rafferty}, D. and {Rawlings}, S. and {Reich}, W. and {de Reijer}, J.-P. and {Reitsma}, J. and {Renting}, G.~A. and {Riemers}, P. and {Rol}, E. and {Romein}, J.~W. and {Roosjen}, J. and {Ruiter}, M. and {Scaife}, A. and {van der Schaaf}, K. and {Scheers}, B. and {Schellart}, P. and {Schoenmakers}, A. and {Schoonderbeek}, G. and {Serylak}, M. and {Shulevski}, A. and {Sluman}, J. and {Smirnov}, O. and {Sobey}, C. and {Spreeuw}, H. and {Steinmetz}, M. and {Sterks}, C.~G.~M. and {Stiepel}, H.-J. and {Stuurwold}, K. and {Tagger}, M. and {Tang}, Y. and {Tasse}, C. and {Thomas}, I. and {Thoudam}, S. and {Toribio}, M.~C. and {van der Tol}, B. and {Usov}, O. and {van Veelen}, M. and {van der Veen}, A.-J. and {ter Veen}, S. and {Verbiest}, J.~P.~W. and {Vermeulen}, R. and {Vermaas}, N. and {Vocks}, C. and {Vogt}, C. and {de Vos}, M. and {van der Wal}, E. and {van Weeren}, R. and {Weggemans}, H. and {Weltevrede}, P. and {White}, S. and {Wijnholds}, S.~J. and {Wilhelmsson}, T. and {Wucknitz}, O. and {Yatawatta}, S. and {Zarka}, P. and {Zensus}, A.},
        title = "{LOFAR: The LOw-Frequency ARray}",
      journal = {\aap},
         year = 2013,
        month = aug,
       volume = {556},
          eid = {A2},
        pages = {A2},
          doi = {10.1051/0004-6361/201220873},
archivePrefix = {arXiv},
       eprint = {1305.3550},
 primaryClass = {astro-ph.IM},
       adsurl = {https://ui.adsabs.harvard.edu/abs/2013A&A...556A...2V}
}

@inproceedings{garakoui2011phased,
  title={Phased-array antenna beam squinting related to frequency dependency of delay circuits},
  author={Garakoui, Seyed Kasra and Klumperink, Eric AM and Nauta, Bram and Van Vliet, Frank E},
  booktitle={2011 41st European Microwave Conference},
  pages={1304--1307},
  year={2011},
  organization={IEEE}
}

@inproceedings{zarka2020low,
  title={The low-frequency radio telescope NenuFAR},
  author={Zarka, Philippe and Denis, Laurent and Tagger, Michel and Girard, Julien and Coffre, Andr{\'e}e and Dumez-Viou, C{\'e}dric and Taffoureau, Christophe and Charrier, Didier and Bondonneau, Louis and Briand, Carine and others},
  booktitle={URSI GASS 2020},
  year={2020}
}

@ARTICLE{2013PASA...30....7T,
       author = {{Tingay}, S.~J. and {Goeke}, R. and {Bowman}, J.~D. and {Emrich}, D. and {Ord}, S.~M. and {Mitchell}, D.~A. and {Morales}, M.~F. and {Booler}, T. and {Crosse}, B. and {Wayth}, R.~B. and {Lonsdale}, C.~J. and {Tremblay}, S. and {Pallot}, D. and {Colegate}, T. and {Wicenec}, A. and {Kudryavtseva}, N. and {Arcus}, W. and {Barnes}, D. and {Bernardi}, G. and {Briggs}, F. and {Burns}, S. and {Bunton}, J.~D. and {Cappallo}, R.~J. and {Corey}, B.~E. and {Deshpande}, A. and {Desouza}, L. and {Gaensler}, B.~M. and {Greenhill}, L.~J. and {Hall}, P.~J. and {Hazelton}, B.~J. and {Herne}, D. and {Hewitt}, J.~N. and {Johnston-Hollitt}, M. and {Kaplan}, D.~L. and {Kasper}, J.~C. and {Kincaid}, B.~B. and {Koenig}, R. and {Kratzenberg}, E. and {Lynch}, M.~J. and {Mckinley}, B. and {Mcwhirter}, S.~R. and {Morgan}, E. and {Oberoi}, D. and {Pathikulangara}, J. and {Prabu}, T. and {Remillard}, R.~A. and {Rogers}, A.~E.~E. and {Roshi}, A. and {Salah}, J.~E. and {Sault}, R.~J. and {Udaya-Shankar}, N. and {Schlagenhaufer}, F. and {Srivani}, K.~S. and {Stevens}, J. and {Subrahmanyan}, R. and {Waterson}, M. and {Webster}, R.~L. and {Whitney}, A.~R. and {Williams}, A. and {Williams}, C.~L. and {Wyithe}, J.~S.~B.},
        title = "{The Murchison Widefield Array: The Square Kilometre Array Precursor at Low Radio Frequencies}",
      journal = {\pasa},
         year = 2013,
        month = jan,
       volume = {30},
          eid = {e007},
        pages = {e007},
          doi = {10.1017/pasa.2012.007},
archivePrefix = {arXiv},
       eprint = {1206.6945},
 primaryClass = {astro-ph.IM},
       adsurl = {https://ui.adsabs.harvard.edu/abs/2013PASA...30....7T}
}

@INPROCEEDINGS{2005ASPC..345..441P,
       author = {{Peterson}, J.~B. and {Pen}, U. and {Wu}, X.},
        title = "{Searching for Early Ionization with the Primeval Structure Telescope}",
    booktitle = {From Clark Lake to the Long Wavelength Array: Bill Erickson's Radio Science},
         year = 2005,
       editor = {{Kassim}, N. and {Perez}, M. and {Junor}, W. and {Henning}, P.},
       series = {Astronomical Society of the Pacific Conference Series},
       volume = {345},
        month = dec,
        pages = {441},
          doi = {10.48550/arXiv.astro-ph/0502029},
archivePrefix = {arXiv},
       eprint = {astro-ph/0502029},
 primaryClass = {astro-ph},
       adsurl = {https://ui.adsabs.harvard.edu/abs/2005ASPC..345..441P}
}

@ARTICLE{2016RAA....16...36H,
       author = {{Huang}, Yan and {Wu}, Xiang-Ping and {Zheng}, Qian and {Gu}, Jun-Hua and {Xu}, Haiguang},
        title = "{The radio environment of the 21 Centimeter Array: RFI detection and mitigation}",
      journal = {Research in Astronomy and Astrophysics},
         year = 2016,
        month = feb,
       volume = {16},
       number = {2},
          eid = {36},
        pages = {36},
          doi = {10.1088/1674-4527/16/2/036},
archivePrefix = {arXiv},
       eprint = {1602.06623},
 primaryClass = {astro-ph.IM},
       adsurl = {https://ui.adsabs.harvard.edu/abs/2016RAA....16...36H}
}

@ARTICLE{2016ApJ...832..190Z,
       author = {{Zheng}, Qian and {Wu}, Xiang-Ping and {Johnston-Hollitt}, Melanie and {Gu}, Jun-hua and {Xu}, Haiguang},
        title = "{Radio Sources in the NCP Region Observed with the 21 Centimeter Array}",
      journal = {\apj},
         year = 2016,
        month = dec,
       volume = {832},
       number = {2},
          eid = {190},
        pages = {190},
          doi = {10.3847/0004-637X/832/2/190},
archivePrefix = {arXiv},
       eprint = {1602.06624},
 primaryClass = {astro-ph.GA},
       adsurl = {https://ui.adsabs.harvard.edu/abs/2016ApJ...832..190Z}
}



\end{document}